\documentclass[conference]{IEEEtran}
\IEEEoverridecommandlockouts
\usepackage{amsmath,amssymb,amsfonts}
\usepackage{graphicx}
\usepackage{mathtools}
\usepackage{textcomp}
\usepackage{xcolor}
\usepackage{pgfplots}

\usepackage[T1]{fontenc}    
\usepackage{xcolor}
\usepackage{url}            
\usepackage{booktabs}       
\usepackage{amsfonts}       
\usepackage{nicefrac}       
\usepackage{microtype}      
\usepackage{subcaption}
\usepackage{float}
\usepackage{amsmath}
\usepackage{algorithm}
\usepackage{algpseudocode}
\usepackage{wrapfig}
\usepackage{tikz}
\usetikzlibrary{positioning}
\usepackage{enumitem}
\usepackage{wrapfig}
\usepackage{marginnote}
\usepackage{natbib}
\usepackage[bookmarks=false]{hyperref}

\newtheorem{definition}{Definition}[section]

\newcommand{\actionspace}{\mathcal{A}}
\newcommand{\noisyscore}{\tilde{q}_a}

\newcommand{\noisyscoreone}{\tilde{q}_1}
\newcommand{\noisyscoretwo}{\tilde{q}_2}

\newcommand{\epsdp}{\epsilon\text{-DP}}

\newcommand{\rsy}{\text{RS}_{\gamma}}
\newcommand{\rnm}{\text{RNM}}
\newcommand{\rnmh}{\text{RNMH}}

\newcommand{\mhom}{\mathcal{M}}
\newcommand{\mhet}{\mathcal{M}'}

\begin{document}

\title{Private Selection with Heterogeneous Sensitivities}

\author{\IEEEauthorblockN{Daniela Antonova}
\IEEEauthorblockA{\textit{Apple} \\
dantonova@apple.com}
\and
\IEEEauthorblockN{Allegra Laro$^*$}\thanks{$^*$Work done while A.L. was at Apple.}
\IEEEauthorblockA{allegralaro@gmail.com}
\and
\IEEEauthorblockN{Audra McMillan}
\IEEEauthorblockA{\textit{Apple} \\
audra.mcmillan@apple.com}
\and
\IEEEauthorblockN{Lorenz Wolf$^\dagger$}\thanks{$^\dagger$Work partly done while L.W. was an intern at Apple.}
\IEEEauthorblockA{\textit{University College London} \\
lorenz.wolf.22@ucl.ac.uk}
}

\maketitle

\begin{abstract}
Differentially private (DP) selection involves choosing a high-scoring candidate from a finite candidate pool, where each score depends on a sensitive dataset. This problem arises naturally in a variety of contexts including model selection, hypothesis testing, and within many DP algorithms. Classical methods, such as Report Noisy Max (RNM) \citep{Dwork_2006Foundations}, assume all candidates' scores are equally sensitive to changes in a single individual's data, but this often isn’t the case. To address this, algorithms like the Generalised Exponential Mechanism (GEM) \citep{raskhodnikova2015efficient} leverage variability in candidate sensitivities. However, we observe that while these algorithms can outperform RNM in some situations, they may underperform in others—they can even perform worse than random selection. In this work, we explore how the distribution of scores and sensitivities impacts DP selection mechanisms. In all settings we study, we find that there exists a mechanism that utilises heterogeneity in the candidate sensitivities that outperforms standard mechanisms like RNM. However, no single mechanism uniformly outperforms RNM. We propose using the correlation between the scores and sensitivities as the basis for deciding which DP selection mechanism to use. Further, we design a slight variant of GEM, \emph{modified GEM} that generally performs well whenever GEM performs poorly. Relying on the correlation heuristic we propose \emph{combined GEM}, which adaptively chooses between GEM and modified GEM and outperforms both in polarised settings.
\end{abstract}

\section{Introduction}

Differentially private (DP) selection is a fundamental task that arises naturally in a variety of contexts including model selection, hypothesis testing and as a subroutine in many algorithms. Given a set of candidates $\actionspace$, a score function $q:\actionspace\times\chi^n\to\mathbb{R}$, and a database $D\in\chi^n$, a differentially private selection algorithm aims to output the candidate with the highest score, $\arg\max_{a\in\actionspace}q(a,D)$, while protecting the privacy of the data subjects whose data is part of $D$. Classical work in private selection, such as Report Noisy Max (RNM) \citep{Dwork_2006Foundations}, assumes that the score function $q(a,\cdot)$ of all selection candidates $a$ is equally sensitive to changes in the dataset $D$. However, in many settings this assumption is false. For example, when performing model selection, more robust candidate models may have lower sensitivity. In personalised recommendations, popular items may have higher sensitivity due to reaching more diverse audiences with varied opinions \citep{kowald2022popularitybiascollaborativefilteringbased}. Inferior items may be universally disliked and hence have low sensitivity \citep{10.1007/s10844-022-00705-9}.

Intuitively, utilizing the fact that some candidates have better than worst-case sensitivity should allow for more accurate differentially private selection mechanisms. Several DP selection mechanisms have been designed with this intuition in mind \citep{Liu_2019PrivateCandidates, raskhodnikova2015efficient}. We find that in a wide variety of settings, mechanisms that utilise heterogeneity in the candidate sensitivities outperform standard mechanisms like RNM. However, we find that no single algorithm uniformly outperforms RNM. That is, for all mechanisms we study, there exists a score function $q$ such that the mechanism performs worse than RNM when instantiated with $q$. In fact, for each mechanism that utilizes heterogeneity in the candidate sensitivities there exists a setting where the mechanism performs worse than random selection. In this work, we aim to design a heuristic that can be used to determine which DP selection mechanism is expected to perform well, and avoid performance worse than random selection.  

When designing differentially private algorithms, it is natural to think that reducing the amount of noise added \emph{anywhere} in an algorithm will improve utility. A surprising finding of our study is that this is not true. There exist settings where adding more noise than necessary actually \emph{improves} performance. To see this, let us consider perhaps the most popular private selection algorithm Report Noisy Max (RNM) \citep{Dwork_2006Foundations}. Define  
the \emph{candidate-wise sensitivity} of a score function for candidate $a\in \actionspace$ to be
\begin{equation}\label{def:candidate_sensitivity}\Delta_a = \underset{D_1,D_2}{\max}|q(a,D_1)-q(a,D_2)|\end{equation}
where the maximum is over all pairs ($D_1$, $D_2$) of datasets that differ on the data of a single individual, and the overall \emph{sensitivity} of the score function to be \begin{equation}\label{def:sensitivity}\Delta=\max_{a\in\actionspace}\Delta_a.\end{equation} Then RNM is defined by 
    \begin{equation}\label{RNM} M(D)= \underset{a\in \actionspace}{\operatorname{argmax}} \; \{q(a,D) + z_a\}, 
    \text{ where } z_a\sim \text{Exp}(\epsilon/2\Delta),
    \end{equation} where $\epsilon$ is the privacy parameter and $\text{Exp}(\epsilon/2\Delta)$ is the exponential distribution with mean $2\Delta/\epsilon$. RNM adds the same amount of noise to the score of every candidate. Consider an alternative algorithm, RNMH\footnote{While RNMH is the natural extension of RNM to include heterogeneous sensitivities, it is not differentially private (see Section~\ref{RNMHnotDP} for details).}, where the amount of noise added is proportional to the candidate-wise sensitivity of the candidate \begin{equation}\label{naivehetRNM}M(D) = \arg\max_{a\in\actionspace}\{q(a,D) + z_a\}, \text{ where } z_a\sim \text{Exp}(\epsilon/2\Delta_a).\end{equation} Since RNMH adds strictly less noise to the scores of some candidates, we might assume that RNMH always outperforms RNM. This is not the case. To see this, let us analyse the behaviour of these two algorithms on a specific example. Suppose we have $k$ candidates, $\actionspace=\{a_1, \cdots, a_k\}$, score function $q$ and database $D$. Suppose there exists $\Delta_1, \Delta_2, q_1$ and $q_2$ such that on the first half of the candidates ($i\le k/2$) $\Delta_{a_i}=\Delta_1$ and $q(a_i,D)=q_1$ and  on the remaining candidates $\Delta_{a_i}=\Delta_2$ and $q(a_i,D)=q_2$. Assume $q_1<q_2$. Using the notation from eqn~\eqref{naivehetRNM}, RNMH outputs one of the higher scoring candidates whenever \begin{equation}\label{RNMHbehaviour}\max_{i=1,\cdots,k/2}z_{a_i}-\max_{i=k/2+1,\cdots,k}z_{a_i}\le q_2-q_1,\end{equation} where $z_{a_i}\sim \text{Exp}(\epsilon/2\Delta_1)$ if $i=1,\cdots,k/2$ and $z_{a_i}\sim \text{Exp}(\epsilon/2\Delta_2)$ otherwise. If $\Delta_1>\Delta_2$ then $\max_{i=1,\cdots,k/2}z_{a_i}$ stochastically dominates $\max_{i=k/2+1,\cdots,k}z_{a_i}$ (that is, it is more likely to output higher values), so the expectation of the LHS of eqn~\eqref{RNMHbehaviour} is positive, grows linearly with $\Delta_1-\Delta_2$, and grows roughly logarithmically with $k$\footnote{The expectation of the LHS is $H_{k/2}(\Delta_1-\Delta_2)$ (where $H_{k/2}$ is the $k/2$th harmonic number).}. If either $k$ or $\Delta_1-\Delta_2$ is large, then the probability of RNMH outputting a higher scoring candidate may be less than 50\%. This is never the case for RNM. Conversely, when $\Delta_1<\Delta_2$, we have the opposite effect and RNMH selects a high scoring candidate more frequently than RNM. 
    
\begin{figure*}[htbp]
        \centering
        \captionsetup{justification=centering}        \begin{subfigure}[b]{\textwidth}
            \begin{subfigure}[b]{0.32\textwidth}
                \includegraphics[width=.9\textwidth, trim={0 0 0 1.2cm},clip]{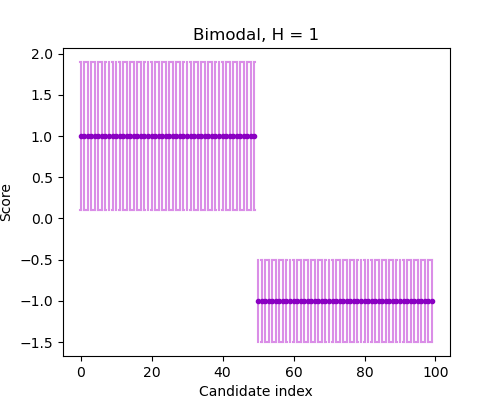}
            \end{subfigure}
            \begin{subfigure}[b]{0.32\textwidth}
                \includegraphics[width=.9\textwidth, trim={0 0 0 1.2cm},clip]{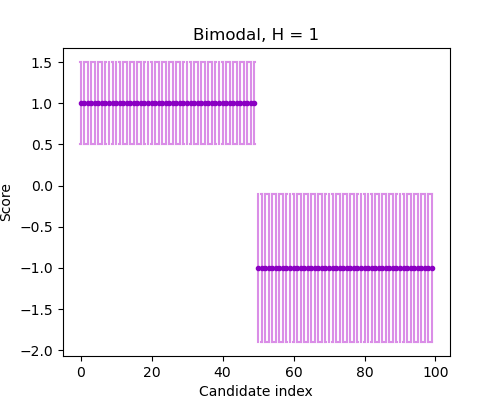}
            \end{subfigure}
            \begin{subfigure}[b]{0.32\textwidth}
        \includegraphics[width=.9\textwidth, trim={0 0 0 1.2cm},clip]{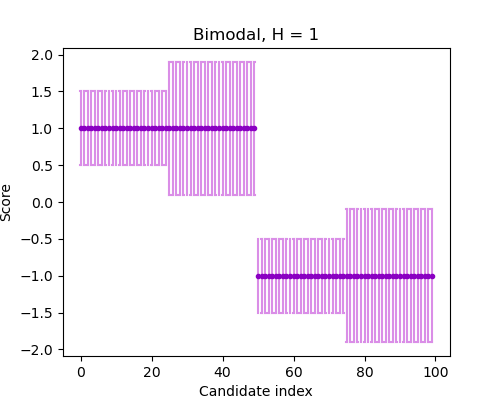}
            \end{subfigure}
        \end{subfigure}
        \begin{subfigure}[b]{\textwidth}
            \begin{subfigure}[b]{0.32\textwidth}
                \includegraphics[width=.83\textwidth]
                {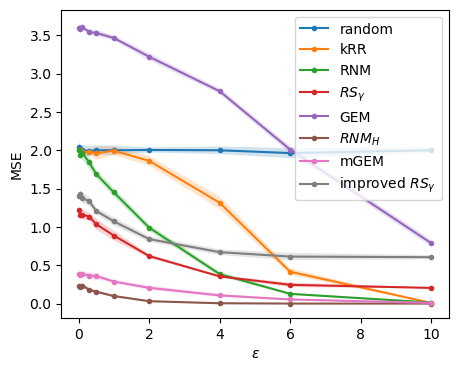}       
                \caption{Scenario 1: high scores = 1, sensitivities = $1.8$; low scores = -1, sensitivities = $1$. }
                \label{fig:scenario_1}
            \end{subfigure}
            \begin{subfigure}[b]{0.32\textwidth}
                \includegraphics[width=.83\textwidth]{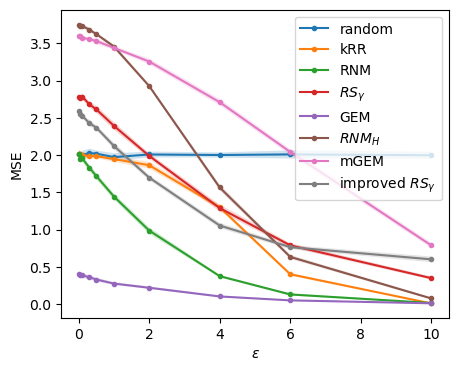}
                \caption{Scenario 2: same scores, but the higher left score group now has small sensitivities.}
                \label{fig:scenario_2}
            \end{subfigure}
            \begin{subfigure}[b]{0.32\textwidth}
                \includegraphics[width=.83\textwidth]{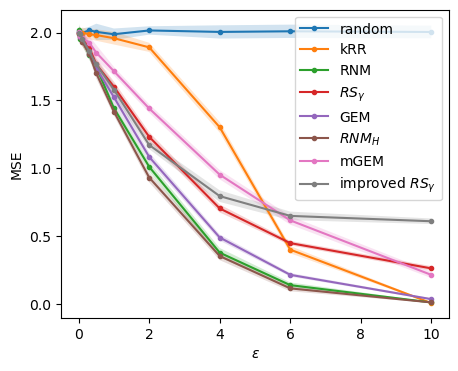}
                \caption{Scenario 3: same scores, but groups have equal share of large and small sensitivities.}
                \label{fig:scenario_3}
            \end{subfigure}       
        \end{subfigure}   
    \caption{Analysis of how the distribution of scores and candidate-wise sensitivities affects the relative  performance of selection algorithms, in three simple scenarios. The figures in the top row show each candidate's scores (dark purple dot) and sensitivities (light purple vertical line). The figures in the second row show the performance (in mean squared error relative to the best candidate) of different private selection algorithms as a function of the privacy parameter $\epsilon$.}
    \label{fig:scenarios_bimodal_alt}
\end{figure*}

    We find that the intuition from the previous example extends to other DP selection mechanisms. That is, the correlation between the candidate scores and sensitivities is a good heuristic for predicting performance relative to RNM (and random selection) for DP selection algorithms that incorporate heterogeneous sensitivities. There have been two main proposals for differentially  private selection algorithms that are able to utilise heterogeneity; the \emph{generalised exponential mechanism} (which we'll denote GEM) \citep{raskhodnikova2015efficient} and RNM with random stopping (which we'll denote $\rsy$), an instantiation of \citet{Liu_2019PrivateCandidates}. RS$_{\gamma}$ is closely aligned to RNMH but achieves formal $(\epsilon, \delta)$-DP (which RNMH does not). We find in all experiments the behavior of RS$_{\gamma}$ is similar to RNMH; it generally performs well under positive correlation and can perform poorly under negative correlation. GEM modifies the score function to favor low sensitivity candidates, so it performs well under negative correlation but can perform poorly under positive correlation. We also propose a modified version of GEM, which we call mGEM, designed to perform well in settings where GEM performs poorly. mGEM has the same general behavior as $\rsy$ but typically outperforms it. While none of these algorithms uniformly outperform RNM, we find that in most natural settings, \emph{one} of these algorithms outperforms RNM. 
    In Figure~\ref{fig:scenarios_bimodal_alt} we examine the relative performance of these algorithms in a setting similar to that described in the previous paragraph. We can see that indeed the expected pattern continues. We'll discuss these and additional results further in Section~\ref{syntheticdata}.

Our contributions are as follows:
\begin{itemize}
    \item We design a variant of GEM, which we call mGEM. While GEM is designed to perform well when the optimal candidate has low sensitivity (the negative correlation case), mGEM performs well when the optimal candidate has high sensitivity. We find that in many natural selection problems (i.e. natural distributions of scores and sensitivities), one of either GEM or mGEM is the optimal DP-selection mechanism, or close to the optimal. In particular, one of these algorithms typically outperforms RNM.
    \item We propose and justify the use of the correlation between the candidate scores and sensitivities as a heuristic to decide which DP-selection mechanism to use. While there exist exceptional cases, this heuristic works well in a variety of settings. We test our heuristic on a variety of synthetic selection problems where we use the Spearman's rank coefficient as our correlation measure. We find that the heuristic works well across a variety of distributions.
    \item We propose \emph{combined GEM}, leveraging the correlation heuristic to adaptively privately choose between mechanisms.
    \item We test our heuristic on three real-world selection problems. In all three problems, we find positive correlation between the scores and sensitivities. We see that, as predicted, our proposed algorithm mGEM is the best performing DP selection mechanism in all three problems. See Figure~\ref{realdata}.
    \item Finally, to further explore the DP selection mechanisms in real-world settings we consider their performance in the context of online bandit problems, where DP selection mechanism often appear as subroutines. We demonstrate that a) private selection can have intrinsic value over non-private selection in an adversarial setting and b) that the previously-observed differences in algorithm performance hold in the online-learning setting.
\end{itemize}

\section{Related work}

The problem of DP selection has been studied extensively in the literature. Most popular DP selection mechanisms, e.g. the Exponential Mechanism \citep{McSherryT07}, the Permute and Flip mechanism \citep{McKenna_2020permute}, and Report Noisy Max \citep{Dwork_2006Foundations} do not utilise heterogeneity in the candidate scores. Several other private selection mechanisms have been proposed with varying assumptions and privacy guarantees \citep{Chaudhuri2014largemargin, Minami2016Differential}.

While most DP selection algorithms focus on homogeneous sensitivities, DP Selection with heterogeneous sensitivities has also been investigated. \citet{raskhodnikova2015efficient} proposed the Generalised Exponential Mechanism to utilise varying sensitivities among candidates. They prove an upper bound on the utility in terms of the sensitivity of the optimal candidate. Motivated by the application of DP hyperparameter-tuning, \citet{Liu_2019PrivateCandidates} proposed a Random Stopping and Private Thresholding algorithm that only requires that the noisy score function is differentially private (i.e. in particular, it allows for the noisy score functions to be tailored to the heterogeneous sensitivities). The authors show  that running an $(\epsilon,0)$-DP algorithm  a random number of times following a geometric distribution, and returning the best output obtained from these runs is $(3\epsilon,0)$-DP (the authors also propose algorithms satisfying $(\epsilon,\delta)$-DP) \citep{Liu_2019PrivateCandidates}. \citet{papernot2022hyperparameter} build on \citet{Liu_2019PrivateCandidates} to obtain results for Rényi DP and show that the privacy guarantee can be retained when geometric distribution is replaced with other distributions. Importantly, by replacing the geometric distribution with the logarithmic distribution, the authors obtain an algorithm satisfying $(2\epsilon,0)$-DP. \citet{koskela2024privacyprofilesprivateselection} further build on the results of \citet{papernot2022hyperparameter} focusing on results for $(\epsilon, \delta)$-DP.

\section{Background}

\subsection{Differential Privacy}
Differential Privacy is often called the gold standard of formal privacy guarantees in machine learning and data science. In this work, we are concerned with discrete outputs, so we will use the following definition:

\begin{definition}[Differential Privacy \citep{Dwork_2006Foundations}]
    Given a privacy loss budget $\epsilon > 0$ and $\delta\in[0,1]$, a randomised algorithm $M:\chi^n\to\actionspace$ satisfies $(\epsilon, \delta)$-differential privacy if, for any set of possible outputs $S\subset \actionspace$ and all pairs of datasets $D_1,D_2$ that differ on the addition or removal of the record(s) of one person, the following holds:
$$Pr[M(D_1) \in S] \leq e^{\epsilon} Pr[M(D_2) \in S]+\delta$$
where the probability is over the randomness of $M$. 
\end{definition}

When $\delta=0$, we will refer to the algorithm as $\epsilon$-differentially private or as satisfying pure differential privacy. If $\delta>0$ then we refer to $M$ as approximately differentially private. In the following subsections, we focus on pure $\epsilon$-DP, unless stated otherwise, where $\epsilon > 0$.
When needed, we will use $\chi$ to represent the data universe, and $\chi^n$ to denote the set of all databases of size $n$. Unless specified otherwise, we will always refer to databases where each person has a single data point/record. Differentially private algorithms typically add carefully calibrated noise to the computation to achieve the privacy guarantee.

\subsection{Private Selection} 
Private selection is the task of selecting a candidate from a list of scored candidates while preserving the privacy of the scores or the data used to compute them. It is a fundamental and well-studied problem in the differential privacy literature \citep{expmech}. Candidate scores are computed using a score function, which represents the utility of an item if it were to be selected. Formally,
given a set of candidates $\actionspace$ and the set of all databases $\chi^n$, a score function takes as input a candidate and a dataset and outputs a real-valued score, i.e. $q:\actionspace\times\chi^n\to\mathbb{R}$.
Given a score function $q$, the selection procedure estimates the item $a\in\actionspace$ with the highest score on the given dataset $D\in\chi^n$, i.e. $\arg\max_{a\in\actionspace} q(a,D)$.

Report Noisy Max (RNM), defined in eqn~\eqref{RNM}, is one of the most popular private selection algorithms. Given any $\epsilon>0$, RNM is $\epsdp$. RNM can be defined with a variety of noise distributions in place of exponential noise including Laplace and Gumbel distributions. We focus on RNM with exponential noise throughout this paper since it has been shown its expected utility dominates the other two noise distributions \citep{McKenna_2020permute, ding2021permuteandflipRNM}. Unless specified otherwise, we will use RNM to refer to RNM with exponential noise. 

A key limitation of RNM is that it adds the same amount of noise to the score of each candidate and is thus unable to utilise this heterogeneity to improve performance. While most of the work on private selection has focused on the homogeneous sensitivities setting (where each candidate has the same variability), there have been two main proposals for selection algorithms that are able to utilise heterogeneity; the \emph{generalised exponential mechanism} \citep{raskhodnikova2015efficient} and RNM with random stopping \citep{Liu_2019PrivateCandidates, papernot2022hyperparameter}. We will give more details on these algorithms in the following sections. 

Finally, we mention one other private selection approach, k-randomised response (kRR) \citep{Warner:1965}. Given $k$ candidates ($|\actionspace|=k$), this algorithm outputs the optimal candidate with probability $e^{\epsilon}/(e^{\epsilon}+k-1)$ and any other candidate with probability $1/(e^{\epsilon}+k-1)$. Unlike RNM, this algorithm does not prioritise outputting candidates that have high scores but are not the optimal. However, it does perform well when $\epsilon$ is large since it downweights \emph{all} candidates except the optimal candidate.

\section{Private Selection with Heterogeneous Sensitivities}
\label{sub:policies}

In this section, we discuss three DP selection algorithms that utilise heterogeneous sensitivities of the candidate scores. Firstly, we will discuss the two existing DP selection algorithms: RS$_{\gamma}$  \citep{Liu_2019PrivateCandidates, papernot2022hyperparameter}, and GEM \citep{raskhodnikova2015efficient}. Finally, we'll introduce a modified version of the Generalised Exponential Mechanism (\emph{GEM}). While GEM is designed to exploit heterogeneous sensitivities when the sensitivity of the optimal candidate is smaller than the the maximum sensitivity $\Delta$, our new version \emph{modified GEM} (mGEM), is designed to perform well when the lower scoring candidates have low sensitivity. 

\subsection{A Naive Extension of RNM with Heterogeneous Sensitivities is Not Differentially Private}\label{RNMHnotDP}

Firstly, let us briefly motivate the introduction of the RS$_{\gamma}$, GEM and mGEM.
It is tempting to incorporate the heterogeneous sensitivities by running RNM as is, but with noise scaled by the candidate-wise sensitivities resulting in RNMH as described in eqn~\eqref{naivehetRNM}. Unfortunately, this algorithm is not $\epsilon$-differentially-private in general. In fact, there exist pairs of sensitivities $\Delta_1$ and $\Delta_2$ such that this algorithm is not $\epsilon'$-DP for any $\epsilon'>0$ even when selecting between just two candidates. Consider an example where candidate $1$ has sensitivity 0 and scores $q_1= 0 = q'_1$ in adjacent datasets $D$ and $D'$. Suppose candidate $2$ have sensitivity 1 and score $q_2 = 1/2$ in $D$ and a score $q'_2=-1/2$ in $D'$. Since exponential noise is strictly positive, the probability of outputting candidate $1$ under dataset $D$ is $0$ but under $D'$ it is $1-e^{-\epsilon/4}$, which implies this algorithm is not $\epsilon'$-DP for any $\epsilon'$.

If we use Laplace noise rather than exponential noise in eqn~\ref{naivehetRNM} (so $z_a\sim \text{Lap}(\epsilon/\Delta_a)$) then it is easy to see that the resulting algorithm is $k\epsilon$-DP where $k$ is the number of candidates. In fact, we can show that it is $(k-1)\epsilon$, although this analysis is tight. There exists a sequence of sensitivities $\Delta_1, \cdots, \Delta_k$ such that RNM with Laplace noise and heterogeneous sensitivities is not $\epsilon'$-DP for any $\epsilon'<(k-1)\epsilon$. A proof of this appears in the appendix.

\subsection{RNM with Random Stopping and Heterogeneous Sensitivities  (RS$_{
\gamma}$)}

In the previous section we established that a naive extension of RNM with heterogeneous sensitivities fails to be differentially private in general. In this section, we are going to augment this algorithm with random stopping to produce an algorithm with many of the benefits of RNMH, but with strict privacy guarantees. The general Randomised Stopping algorithm we will use was introduced by \citet{Liu_2019PrivateCandidates} as an algorithm for privately outputting the highest scoring candidate where the score functions themselves are differentially private. We instantiate this algorithm with differentially private score functions that perturb the score with Laplacian noise. Pseudo-code is given in Algorithm~\ref{alg:random_stopping}, we will denote this algorithm by RS$_{\gamma}$. RS$_{\gamma}$ is $\epsilon$-DP for any choice of $\gamma$.

\begin{algorithm}
\caption{RNM with Random Stopping and Heterogeneous Sensitivities (RS$_{
gamma}$)}\label{alg:random_stopping}
\begin{algorithmic}
\State \textbf{Inputs:} budget $\gamma \leq 1$, access to $q(\cdot,D)$.
\State Initialize list $\mathcal{S} = \emptyset$
\For{$j=1 \dots \infty$}
\State Draw $a$ uniformly at random from $\actionspace$.
\State Sample $\tilde{q}_a= q(a,D) + z_a, \text{ where } z_a\sim \text{Lap}(3\Delta_a/\epsilon)$
\State $\mathcal{S} \gets \mathcal{S} \cup \{(a, \tilde{q}_a)\}$
\State With probability $\gamma$ output the highest scored candidate from $\mathcal{S}$ and halt
\EndFor
\end{algorithmic}
\end{algorithm}

For a fixed privacy guarantee, random stopping balances utility and computational cost, with the help of an additional hyper-parameter $\gamma$. Note that when $\gamma$ is small, RS$_{\gamma}$ is similar in spirit to RNMH but adds slightly more noise to the score of each candidate. Thus, we expect RS$_{\gamma}$ to follow a similar pattern of behaviour as RNMH but with slightly worse performance. For a more in-depth discussion of the performance of RS$_{\gamma}$ as $\gamma$ varies, see Appendix~\ref{appendix:two_candidates}.

In \citet{papernot2022hyperparameter}, the authors propose to sample the number of trials from a truncated negative binomial distribution, parameterised by $\gamma\in (0,1)$ and $\eta\in (-1, \infty)$, instead of the Geometric distribution, which results in a mechanism satisfying $(2+\eta)\epsilon$-DP (details in Appendix \ref{app:improved_rs}). We denote this version with $\eta=0$ as \emph{improved $\rsy$} in our experiments.

\subsection{Generalised Exponential Mechanism (GEM)}

The Generalised Exponential Mechanism, proposed in \citet{raskhodnikova2015efficient}, also addresses private selection with heterogeneous sensitivities. It does so by transforming the score function $q$ into a new score function $q'$ such that $q'$ has overall sensitivity at most 1: \begin{align}q'(a, D) &\coloneqq \underset{a'\in\actionspace}{\min}\frac{(q(a,D) - t\Delta_a) - (q(a',D) - t\Delta_{a'})} {\Delta_a + \Delta_{a'}}.
\label{eq:gem}\\
&= \underset{a'\in\actionspace}{\min}\frac{(q(a,D) - q(a',D)) - t(\Delta_a -  \Delta_{a'})} {\Delta_a + \Delta_{a'}}.\label{eq:gemotherformulation}
\end{align}
The normalised scores $q'$ are then input into RNM\footnote{This algorithm is called the Generalised Exponential mechanism because \citet{raskhodnikova2015efficient} originally proposed using the new score function with the exponential mechanism, another standard private selection algorithm. We will use RNM with Exponential noise since it typically outperforms the exponential mechanism \citep{McKenna_2020permute, ding2021permuteandflipRNM}.}. \citet{raskhodnikova2015efficient} set
$t=\frac{2\log(|\actionspace|/\beta)}{\epsilon}$, where $\beta$ controls the probability of a "bad" outcome and thus the error of the algorithm. \citet{raskhodnikova2015efficient} prove an upper bound on the error of the Generalised Exponential Mechanism that is proportional to sensitivity of the optimal candidate. However, they do not explore how the distribution of the sensitivities $\Delta_a$ affects performance. We find that GEM can perform poorly in cases where the sensitivity of the optimal candidate is close to the worst-case sensitivity.

\subsection{Modified Generalised Exponential Mechanism (mGEM)}\label{s:modgemtheory}

In this section we introduce a variant of GEM, mGEM, that is designed for the setting where we have positive correlation between the scores and sensitivities. \citet{raskhodnikova2015efficient} propose to set $t = \frac{2\log(|\actionspace|/\beta)}{\epsilon}$, regardless of the data. They derive this value for $t$ by following the standard utility bound analysis for the exponential mechanism, i.e., that the score of the chosen item $q'_{\hat{1}}$ is at least the score of the best item $q'_{\tilde{1}}$ minus a factor that depends inversely on $\epsilon$:

$$q'_{\hat{1}} \geq q'_{\tilde{1}} - \frac{2 \log(|\actionspace|/\beta)}{\epsilon}$$

By substituting the mapped GEM scores into this inequality, observing that score of the optimal candidate (according to $q'$) is zero, and rearranging, we have that, for all $j$, the score of the chosen candidate $q_{\hat{i}}$ satisfies:

$$ q_{\hat{1}} \geq q_j - \Delta_{\hat{1}} \left(\frac{2 \log(|\actionspace|/\beta)}{\epsilon} - t\right) - \Delta_j \left(\frac{2 \log(|\actionspace|/\beta)}{\epsilon} + t\right)$$

By setting $t=\frac{2\log(|\actionspace|/\beta)}{\epsilon}$, GEM achieves the utility bound \[q_{\hat{1}} \geq \max_j\{q_j - \Delta_j \frac{4 \log(|\actionspace|/\beta)}{\epsilon}\}.\] When there is a negative correlation between $q_j$ and $\Delta_j$, this is a strong guarantee since the error (the difference between the score of the chosen candidate and the optimal candidate) scales, at most, with the sensitivity of the optimal score, which is small. An alternate explanation is that by setting $t$ to be positive, GEM penalizes candidates with high sensitivity, which are correlated with candidates with low scores. In the case of positive correlation, the utility guarantee for GEM is weaker since the sensitivity of the optimal candidate may be large. Further, penalizing candidates with high sensitivity is a poor choice under positive correlation since this penalizes high scoring candidates. Hence, we propose setting $t=-\frac{2\log(|\actionspace|/\beta)}{\epsilon}$, which penalizes candidates with \emph{low} sensitivities. We'll return to this intuition in Section~\ref{syntheticdata}.

\section{Intuition for How the Distribution of Scores and Sensitivities Affects Performance.}

In this section, we build intuition for how the distribution of scores and sensitivities affects performance of the DP selection mechanisms. We will focus on the comparison between GEM and mGEM since, in all scenarios we studied, one of these mechanisms was the best performing (or comparable to the best performing) DP mechanism. We first discuss the simple two-candidate setting and discuss how positive or negative correlation between the scores and sensitivities determines the relative performance of GEM, mGEM, RNM and random selection. We'll then give some intuition for why correlation can generally be used as a heuristic for determining the relative performance even in the multiple candidates setting.

\subsection{The Two Candidate Setting}\label{twocandidates}

Let us begin with a study of the two candidate setting so $\actionspace=\{a_1,a_2\}$. Let us just consider the comparison between GEM and mGEM in this setting. Given a database $D$, denote the scores of $a_1$ and $a_2$ by $q_1=q(a_1,D)$ and $q_2=q(a_2,D)$. Without loss of generality let $q_1 < q_2$. Let $\Delta_1$ and $\Delta_2$ be the candidate-wise sensitivities of $a_1$ and $a_2$, respectively. 

Let us first consider the behavior of GEM in this setting. Let $q'_1$ and $q'_2$ be the modified scores as defined in eqn~\ref{eq:gem} with $t=\frac{2\log(|\mathcal{A}|/\beta)}{\epsilon}$. If $\Delta_1>\Delta_2$, then $q_1-t\Delta_1<q_2-t\Delta_2$, which implies $q_1'=\frac{(q_1-t\Delta_1)-(q_2-t\Delta_2)}{\Delta_1+\Delta_2}<0$ and $q'_2=0$. Therefore, candidate 2 remains the favored outcome. In fact, we expect GEM to favor candidate 2 more heavily than RNM in this setting since candidate 1 is penalised for having a large sensitivity. Conversely, if $\Delta_1<\Delta_2$ then is it possible that $q_1-t\Delta_1>q_2-t\Delta_2$, so $q_1'=0$ and $q'_2=\frac{(q_2-t\Delta_2)-(q_1-t\Delta_1)}{\Delta_1+\Delta_2}<0$. In this setting, candidate 1 is the favored output. Thus, GEM will not only perform worse than RNM in this setting, but will perform worse than random selection since it favors the wrong candidate. This occurs both when there is a large gap in the sensitivities and when $\epsilon$ is small (so $t$ is large).

The mechanism mGEM displays the opposite behavior. Performing the same exercise on the modified scores for mGEM we see that if $\Delta_1<\Delta_2$ then we expect mGEM to favor candidate 2 more heavily than RNM. Conversely, if $\Delta_1>\Delta_2$ then mGEM can perform worse than random selection since it can favor candidate 1.

We find empirically that mGEM and RS$_{\gamma}$ perform similarly. That is, they outperform RNM in the same settings, and perform worse than RNM and random selection in the same settings. Next, we will discuss why RS$_{\gamma}$ behaves this way. We will use RNMH as a proxy for RS$_{\gamma}$ since they are designed to behave similarly and  RNMH is more intuitive to analyse. We will compare RNM and RS$_{\gamma}$ on their probabilities of outputting the lower scoring candidate. That is, for two mechanisms $\mhom$ and $\mhet$, define the metric
    \begin{align*}
        HG(\mhom,\mhet) = Pr[\mhom(q) \neq a^*]- Pr[\mhet(q) \neq a^*],
    \end{align*}
    where $a^*$ is the optimal candidate. Then $HG(\mhom,\mhet)>0$ if and only if $\mhet$ outperforms $\mhom$.

In the case of $\rnm$, scores are noised according to $\noisyscore = q_a + z_a$, where $z_a \sim \text{Exp}(\epsilon/2\Delta)$ and $\Delta$ is the global sensitivity $\Delta=\max(\Delta_1, \Delta_2)$. The probability of outputting the lower scoring candidate is
$$Pr(\noisyscoreone > \noisyscoretwo) = \frac{1}{2} \exp{\left( -\frac{\epsilon}{2\Delta} (q_2-q_1)\right)}.$$ 
In RNMH, $\noisyscore = q_a + z_a$, where $z_a \sim \text{Exp}(\epsilon/2\Delta_a)$ so the probability of outputting the lower scoring candidate is
$$Pr(\noisyscoreone > \noisyscoretwo) = \frac{1}{1+\Delta_2/\Delta_1} \exp{\left( -\frac{\epsilon}{2\Delta_1} (q_2-q_1)\right)}.$$

\begin{figure}[t]
\centering

\includegraphics[width=.45\textwidth]{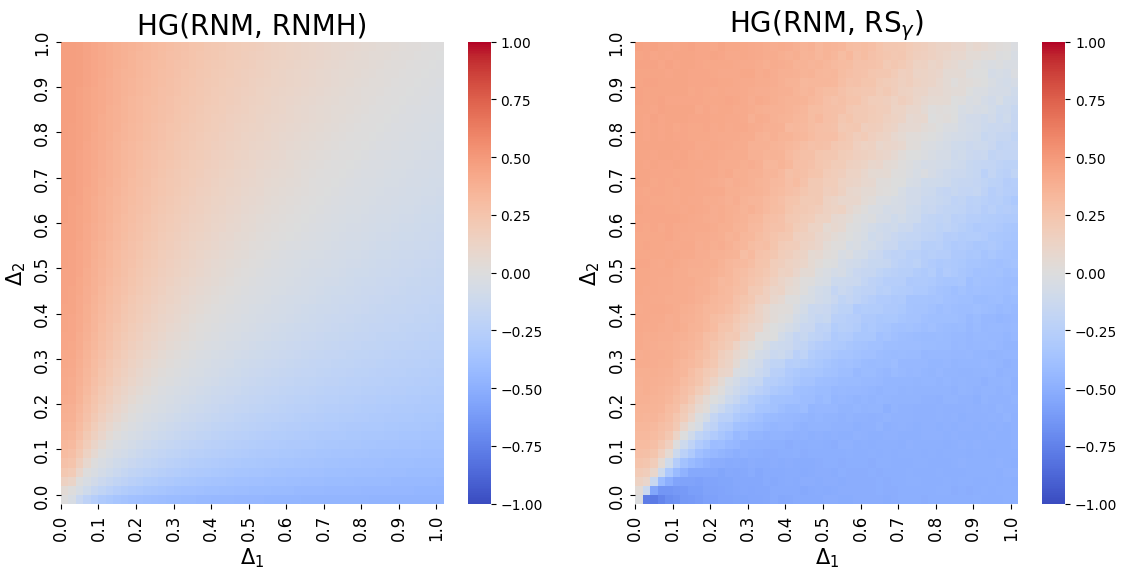}
\caption{Comparing the performance of $\rnm$, RMNH and $\rsy$ for varyiing different sensitivities $\Delta_1$ (on the horizontal axis) and $\Delta_2$ (on the vertical axis). Here $\rsy$ is run with $\gamma=0.01$, $\epsilon=0.1$ and scores $q_1=0, q_2=1$.} 
\label{fig:hg_hetmechs}
\end{figure}

It follows that $HG(\rnm, \rnmh)>0$ for $\Delta_1<\Delta_2$ and $HG(\rnm, \rnmh)<0$ for $\Delta_1>\Delta_2$. That is, when the higher scoring candidate has larger sensitivity RNMH outperforms RNM. Otherwise, RNM performs better. In Figure~\ref{fig:hg_hetmechs}, we compute $HG($RNM, RS$_{\gamma})$ empirically, confirming that it does indeed follow the pattern predicted by analysing RNMH. These observations are supported by additional empirical results in Appendix~\ref{appendix:two_candidates}.

\subsection{The Multiple Candidate Setting}\label{syntheticdata}
\begin{figure}[t]
        \centering
        
        \begin{subfigure}[b]{0.24\textwidth}
\begin{tikzpicture}[scale=0.5]
    \begin{axis}[
        axis lines=middle,
        xlabel={Candidates},  
        ylabel={Scores},
        xlabel style={at={(axis description cs:0.5,-0.1)},anchor=north},  
        ylabel style={at={(axis description cs:-0.1,0.5)},rotate=90,anchor=south},  
        xmin=0, xmax=3,
        ymin=0, ymax=8,
        domain=-3:3,
        samples=100,
        xtick=\empty,   
        ytick=\empty,
        legend pos=north west
    ]

    \addplot[
        ultra thick,
        dotted,
        black,
        restrict y to domain=2:5
    ] coordinates {(0, 2) (3, 5)};
    
    \addplot[
        ultra thick,
        dotted,
        color={rgb,255:red,128; green,0; blue,128}
    ] coordinates {(0, 1.5) (3, 2.5)};
    
    \addplot[
        ultra thick,
        dotted,
        color={rgb,255:red,255; green,105; blue,180}
    ] coordinates {(0, 2.5) (3, 7.5)};

    \addplot[<->,
        ultra thick,
    ] coordinates {(1.5,3.6) (1.5, 4.8)};

    \node at (axis cs:1.7, 4.2) {$t \Delta_a$}; 

    \addplot[<->,
        ultra thick,
    ] coordinates {(1.5,2.2) (1.5, 3.4)};

    \node at (axis cs:1.7, 2.8) {$t \Delta_a$}; 

    \end{axis}
\end{tikzpicture}
\caption{Positive correlation.}
\end{subfigure}
\hfill
        \begin{subfigure}[b]{0.24\textwidth}
\begin{tikzpicture}[scale=0.5]
    \begin{axis}[
        axis lines=middle,
        xlabel={Candidates},  
        ylabel={Scores},
        xlabel style={at={(axis description cs:0.5,-0.1)},anchor=north},  
        ylabel style={at={(axis description cs:-0.1,0.5)},rotate=90,anchor=south},  
        xmin=0, xmax=3,
        ymin=0, ymax=8,
        domain=-3:3,
        samples=100,
        xtick=\empty,   
        ytick=\empty,
        legend pos=north west
    ]

    \addplot[
        ultra thick,
        dotted,
        black,
        restrict y to domain=2:5
    ] coordinates {(0, 3) (3, 5)};
    
    \addplot[
        ultra thick,
        dotted,
        color={rgb,255:red,255; green,105; blue,180}
    ] coordinates {(0, 5.5) (3, 5.5)};
    
    \addplot[
        ultra thick,
        dotted,
        color={rgb,255:red,128; green,0; blue,128}
    ] coordinates {(0, 0.5) (3, 4.5)};

    \addplot[<->,
        ultra thick,
    ] coordinates {(1.5,4.1) (1.5, 5.3)};

    \node at (axis cs:1.7, 4.7) {$t \Delta_a$}; 

    \addplot[<->,
        ultra thick,
    ] coordinates {(1.5,2.7) (1.5, 3.9)};

    \node at (axis cs:1.7, 3.3) {$t \Delta_a$}; 

    \end{axis}
\end{tikzpicture}
\caption{Negative correlation.}
\end{subfigure}
\caption{Analysing the impact of correlation between scores and sensitivities on the behavior of GEM and mGEM. The centre black line is $q_a$, the top (pink) line is the function used to reorder candidates for mGEM, and the bottom purple line is the function used to reorder candidates for GEM.}\label{pictureofGEMscores}
\end{figure}
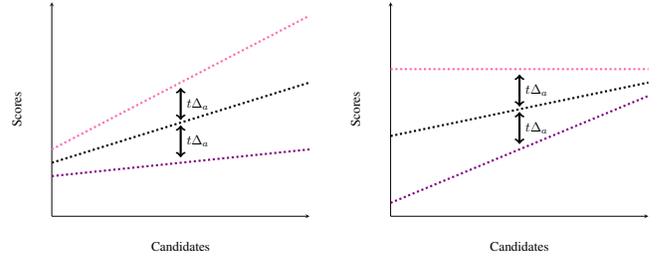

Now let us turn to the general multiple candidates setting. To gain some intuition, let us look at some extreme cases of correlation between the scores and sensitivities. Let $t=\frac{2\log(|\mathcal{A}|/\beta)}{\epsilon}$. Observe from eqn~\eqref{eq:gem} that, if we ignore the normalisation, GEM reorders the candidates according to $q(a,D)-t\Delta_a$ while mGEM reorders the candidates according to $q(a,D)+t\Delta_a$. In Figure~\ref{pictureofGEMscores} we visualise what these score functions look like in a simple case of positive and negative correlation between the scores and sensitivities. Suppose the candidates are ordered according to $q(\cdot,D)$ and indexed on the horizontal axis. The black line is $q(a, D)$. The top pink line is $q(a,D)+t\Delta_a$, the function used to reorder mGEM, and the bottom purple line is $q(a,D)-t\Delta_a$, the function used to reorder GEM. Notice that in the case of positive correlation, the pink line (mGEM) has a more positive slope than the black line, further biasing it towards candidates with high score $q(a,D)$. Thus, we expect (and observe experimentally) that mGEM will outperform RNM in this setting. The purple line (GEM) has a less positive slope and if $t$ is large enough, the purple line can even become decreasing. If the line becomes decreasing then GEM will actually favor \emph{lower} scoring candidates, causing it to perform worse than random selection. We see the opposite behavior in the negative correlation case. In this case, we expect (and observe experimentally), based on the slopes of the corresponding lines, that RNM with outperform mGEM but be outperformed by GEM. We reach similar conclusions via utility bound analysis; see Section \ref{a:ubounds} for details.

\begin{figure}[t]
        \centering
        \captionsetup{justification=centering}
  \begin{subfigure}[b]{0.5\textwidth}
            \begin{subfigure}[b]{0.49\textwidth}              
                \includegraphics[width=.9\textwidth, trim={0 0 0 .71cm},clip]{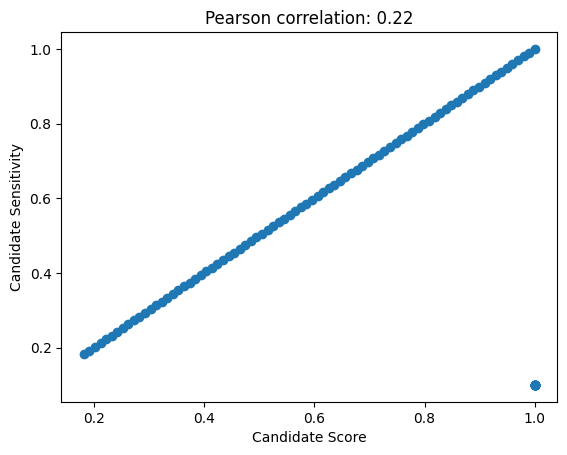}
                \caption{Scenario 1: \\Pearson Correlation: $0.22$\\
                        Spearman Correlation: $0.12$}
                \label{fig:scenario11}
            \end{subfigure}
            \begin{subfigure}[b]{0.49\textwidth}              
                \includegraphics[width=.9\textwidth, trim={0 0 0 .71cm},clip]{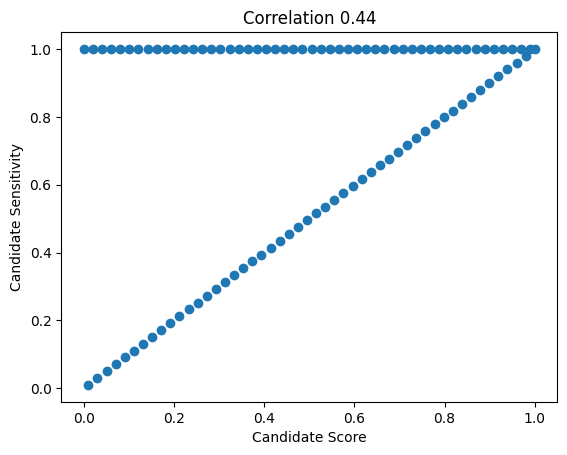}
                \caption{Scenario 2: \\Pearson Correlation: $0.44$\\
                        Spearman Correlation: $0.27$}
                \label{fig:scenario21}
            \end{subfigure}
        \end{subfigure}
        \begin{subfigure}[b]{.5\textwidth}
            \begin{subfigure}[b]{0.49\textwidth}
                \includegraphics[width=.9\textwidth,clip]{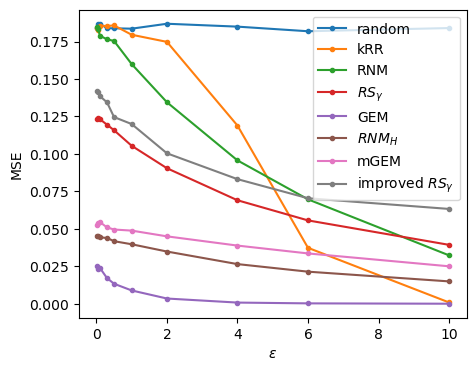}
                \caption{Performance of DP selection mechanisms on Scenario 1}
                \label{fig:scenario1perf}
            \end{subfigure}
            \begin{subfigure}[b]{0.49\textwidth}
                \includegraphics[width=.9\textwidth,clip]{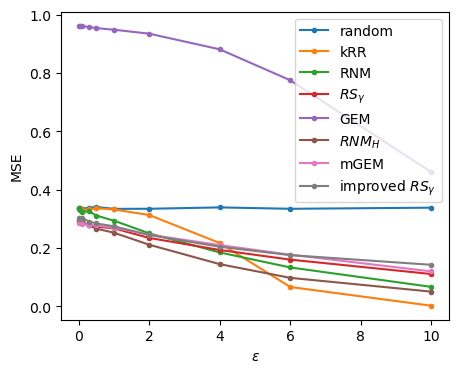}
                \caption{Performance of DP selection mechanisms on Scenario 2}
                \label{fig:scenario2perf}
            \end{subfigure}
    \end{subfigure}
        \begin{subfigure}[b]{0.5\textwidth}
            \begin{subfigure}[b]{0.49\textwidth}
                \includegraphics[width=.9\textwidth, trim={0 0 0 .71cm},clip]{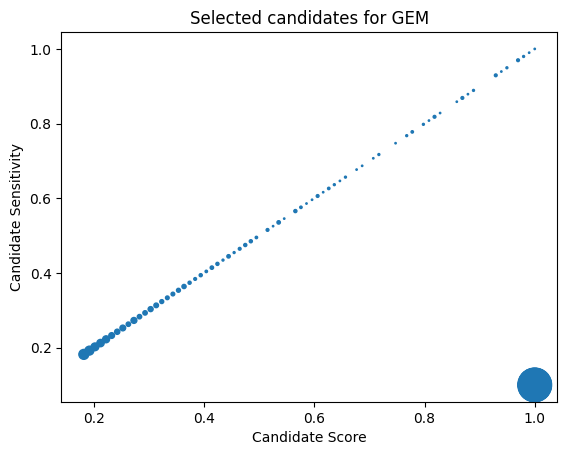}
                \caption{Selected candidates for GEM on Scenario 1}
                \label{fig:scenario1GEM}
            \end{subfigure}
                        \begin{subfigure}[b]{0.49\textwidth}
                \centering
                \captionsetup{justification=centering}
                \includegraphics[width=.9\textwidth, trim={0 0 0 .7cm},clip]{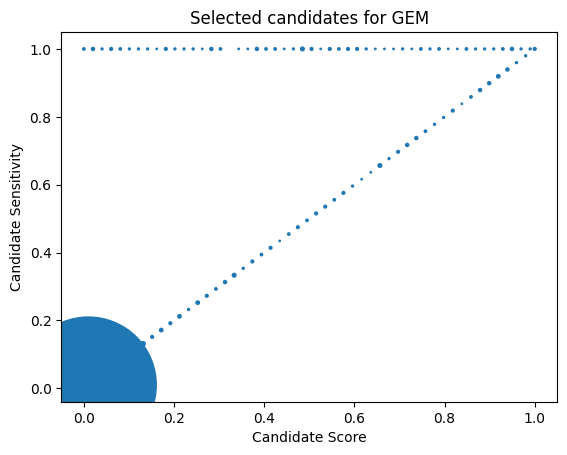}
                \caption{Selected candidates for GEM on Scenario 2}
                \label{fig:weights_gem}
            \end{subfigure}
                \label{fig:scenario2GEM}
        \end{subfigure}
        \begin{subfigure}[b]{.5\textwidth}
            \begin{subfigure}[b]{0.49\textwidth}
                \centering
                \includegraphics[width=.9\textwidth, trim={0 0 0 .71cm},clip]{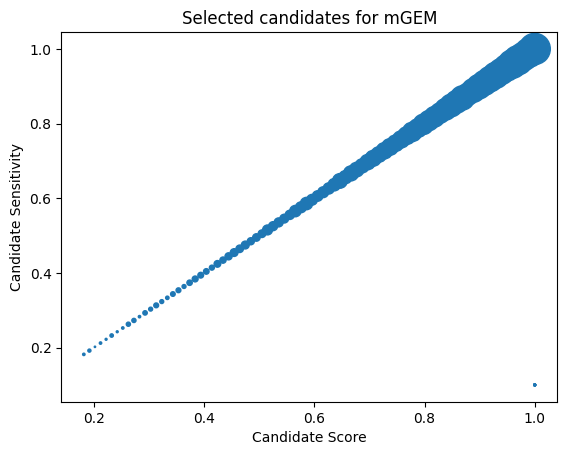}
                \caption{Selected candidates for mGEM on Scenario 1}
                \label{fig:scenario1mGEM}
            \end{subfigure}
            \begin{subfigure}[b]{0.49\textwidth}
                \centering
                \captionsetup{justification=centering}
                \includegraphics[width=.9\textwidth, trim={0 0 0 .7cm},clip]{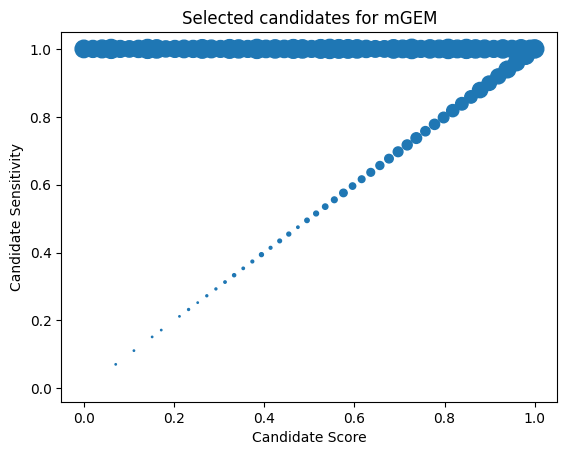}
                \caption{Selected candidates for mGEM on Scenario 2}
                \label{fig:scenario2mGEM}
            \end{subfigure}
        \end{subfigure}   
        
    \caption{Figure~\ref{fig:scenario11} and~\ref{fig:scenario21} in the top row show the sensitivities plotted against scores (Scenario 1 has $18$ candidates overlaid with score $1$ and sensitivity $0.1$). Figures~\ref{fig:scenario1perf} and~\ref{fig:scenario2perf} show mechanism performance in terms of MSE for different values of the privacy parameter $\epsilon$. In Figures~\ref{fig:scenario1GEM}-\ref{fig:scenario2mGEM}, candidates are weighted by how likely the specified mechanism is to choose that candidate (with $\epsilon=0.05$). }
    \label{fig:correlation_scenarios}
\end{figure}

Of course, in most real selection problems, the sensitivity of candidates will not strictly increase or decrease with the candidate scores. In the next section we will test our heuristic on more complex distributions of scores and sensitivities. For the remainder of this section, let us briefly discuss some exceptional cases. In Figure~\ref{fig:correlation_scenarios}, we explore the behavior of GEM and mGEM on two more interesting scenarios. According to both the Pearson and Spearman correlation measures, both of these scenarios are instances of positive correlation. In Scenario 1, as expected mGEM outperforms RNM. While mGEM rarely selects the optimal candidate in this scenario, this doesn't affect its MSE much since it selects near optimal candidates with high sensitivity. However, GEM also performs well in this scenario. This is because despite the overall positive correlation, the optimal candidate has low sensitivity resulting in GEM having low error in this scenario. 

In Scenario 2 in Figure~\ref{fig:correlation_scenarios}, we see that GEM does indeed perform poorly, as expected. It consistently selects low scoring but low sensitivity candidates. However, while mGEM performs reasonably well in this scenario, it is outperformed by RNM at higher values of $\epsilon$. This is because low scoring but high sensitivity candidates are favored by mGEM in this scenario. Thus, while correlation serves as a useful heuristic for assessing the relative performance of differential privacy (DP) selection mechanisms, it does not capture the complete picture in certain exceptional cases.
Nevertheless, such exceptional cases do not preclude using correlation as a heuristic in practice, where average behavior may be important.

\section{Analysing Correlation as a Heuristic}

In this section, we evaluate the DP selection mechanisms on synthetically generated distributions of scores and sensitivities. We create a number of different scenarios representing different distributions of the set $\{(q(a,D), \Delta_a)\}$. 

Of course, these scenarios do not provide an exhaustive list, but they are do provide intuition for situations that may arise in practice. 
Since in this section we are studying the performance of the algorithms on a specific dataset $D$, we will drop references to $D$ in the notation and use $q_a$ to denote $q(a,D)$. We include k-randomised response (denoted kRR) and the algorithm that outputs a candidate drawn uniformly at random (denoted random) as further comparison points. We also include RNMH for reference, even though it does not satisfy DP guarantees. These experiments expand on the intuition we gained in the previous section. In particular, we confirm that our proposed heuristic of using the correlation between scores and sensitivities still works well on more complex distributions of candidate scores and sensitivities. 

\paragraph{Experimental Design}

In each scenario we study, for each candidate $a$, we define a distribution $P_a$. In each trial, we will sample a new score for each candidate $a$ from $P_a$. We include randomness in the candidate scores between trials to mimic the fact that in the real world, a domain expert may have an expectation of the relationship between the scores and sensitivities but any particular instantiation will have some variability. The sensitivity for candidate $a$ is constant across all trials. To define $\Delta_a$, we first sample $N$ sets of scores $\{q_a\}_{a\in\mathcal{A}}$, one for each trial. The sensitivity $\Delta_a$ is then defined to be the difference between the $10th$ and $90th$ quantiles of the $N$ scores sampled from $P_a$. Any scores $q_a$ that are not contained within these quantiles are subsequently clipped to this range. 

To evaluate a given selection mechanism in a particular scenario (defined by the distributions $P_a$) we generate $N$ sets 
$\{(q_a, \Delta_a)\}_{a\in\mathcal{A}}$ as described above. Each of these $N$ datasets corresponds to a single trial, where we define the error for that trial as the difference between the score of the selected candidate and the score of the optimal candidate (which may vary between trials). The overall error is defined as the mean squared error (MSE) over all $N$ trials. Note that we compute the MSE on the clipped scores. Clipping is often performed in practice to ensure finite sensitivity.

\begin{figure*}[t]
        \centering
        \captionsetup{justification=centering}        \begin{subfigure}[b]{\textwidth}
            \begin{subfigure}[b]{0.33\textwidth}
                \includegraphics[width=.9\textwidth, trim={0 0 0 1.2cm},clip]{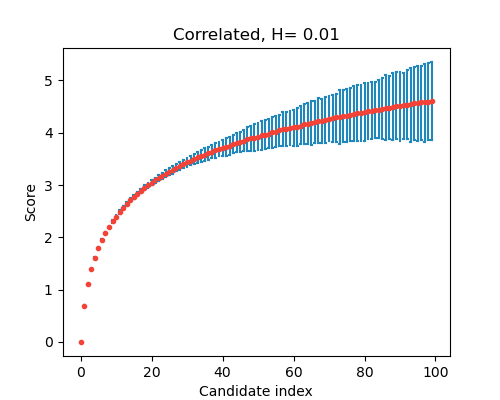}
            \end{subfigure}
            \begin{subfigure}[b]{0.33\textwidth}
                \includegraphics[width=.9\textwidth, trim={0 0 0 1.2cm},clip]{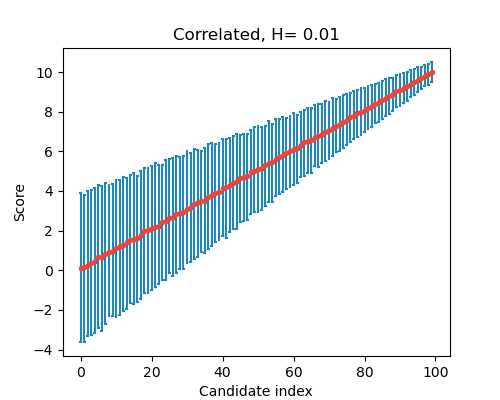}
            \end{subfigure}
            \begin{subfigure}[b]{0.33\textwidth}
                \includegraphics[width=.9\textwidth, trim={0 0 0 1.2cm},clip]{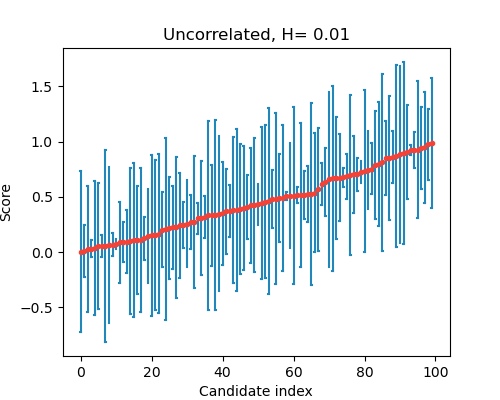}
            \end{subfigure}
        \end{subfigure}
        \begin{subfigure}[b]{\textwidth}
            \begin{subfigure}[b]{0.33\textwidth}
                \includegraphics[width=.83\textwidth]{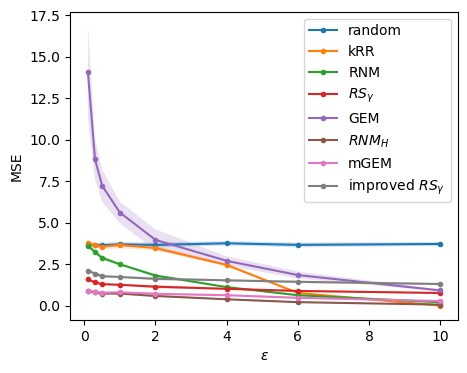}
                \caption{Scenario 4}
                \label{fig:scenario4}
            \end{subfigure}
            \begin{subfigure}[b]{0.33\textwidth}
                \includegraphics[width=.83\textwidth]{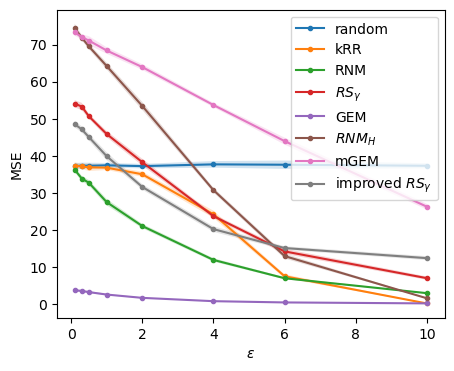}
                \caption{Scenario 5\\\hspace{\textwidth} }
                \label{fig:scenario5}
            \end{subfigure}
            \begin{subfigure}[b]{0.33\textwidth}
                \includegraphics[width=.83\textwidth]{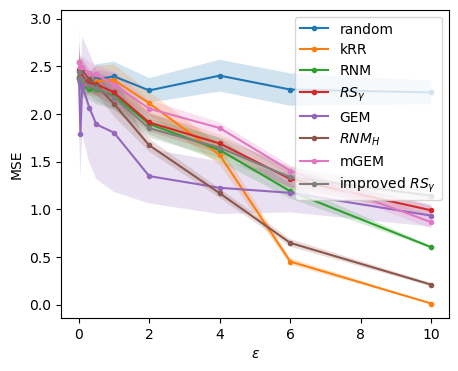}
                \caption{Scenario 6}
                \label{fig:scenario6}
            \end{subfigure}
        \end{subfigure}   
    \caption{Analysis of how the distribution of scores and candidate-wise sensitivities affects the relative  performance of algorithms in three slightly more realistic scenarios, with some randomness in the scores. The figures in the top row show the distributions according to which scores and sensitivities are obtained. Each \emph{mean} score is shown as a red dot and each sensitivity as a blue vertical line. The figures in the second row show mechanism performance in terms of MSE for different values of the privacy parameter $\epsilon$.}   
    \label{fig:second_scenarios}
\end{figure*}

\begin{figure*}[htbp]
    \centering
    \begin{subfigure}[b]{\textwidth}
        \begin{subfigure}[b]{0.16\textwidth}
            \includegraphics[width=\textwidth]{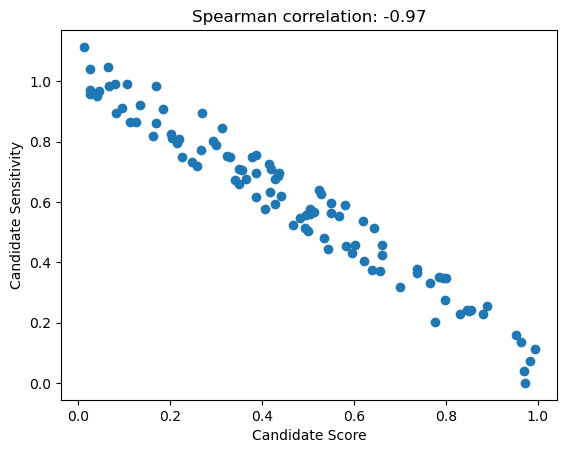}
       \label{fig:corr_linear_0}
        \end{subfigure}
        \hfill
        \begin{subfigure}[b]{0.16\textwidth}
            \includegraphics[width=\textwidth]{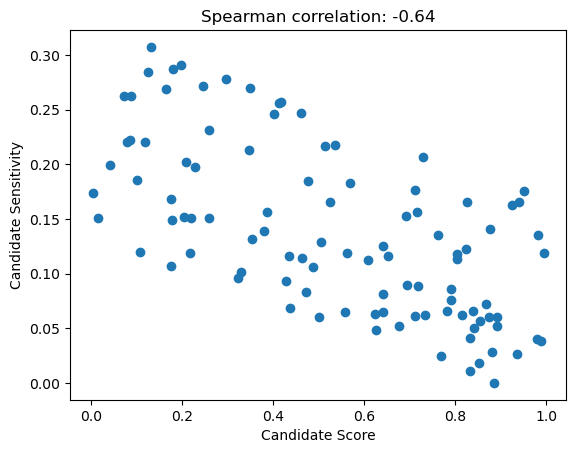}
        \label{fig:corr_linear_1}
        \end{subfigure}
        \hfill
        \begin{subfigure}[b]{0.16\textwidth}
            \includegraphics[width=\textwidth]{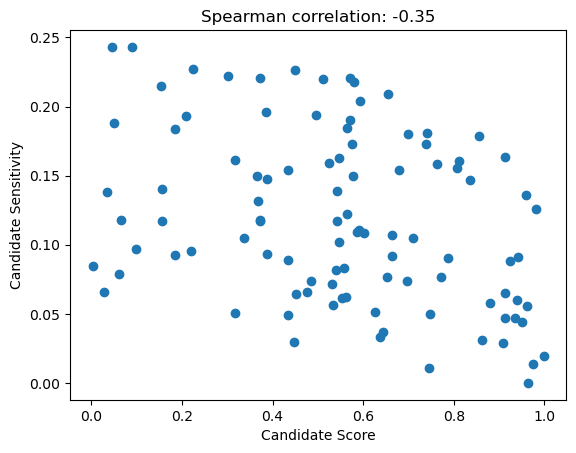}
        \label{fig:corr_linear_2}
        \end{subfigure}
        \hfill
        \begin{subfigure}[b]{0.16\textwidth}
            \includegraphics[width=\textwidth]{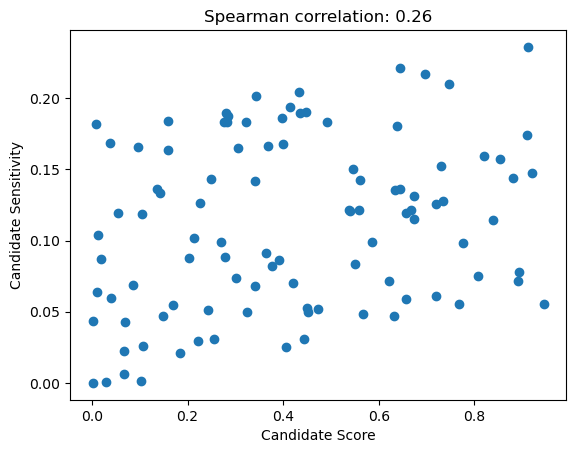}
        \label{fig:corr_linear_4}
        \end{subfigure}
        \hfill
        \begin{subfigure}[b]{0.16\textwidth}
            \includegraphics[width=\textwidth]{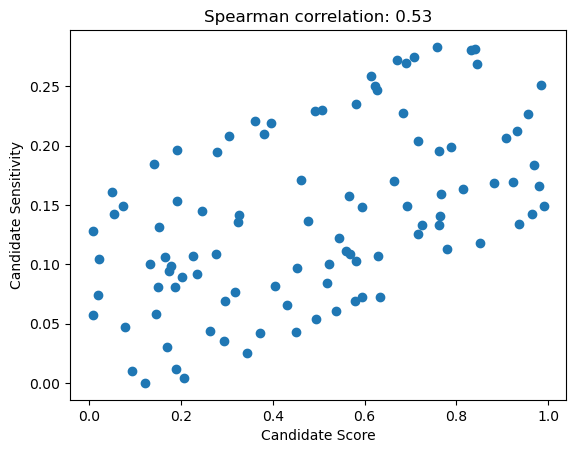}
        \label{fig:corr_linear_5}
        \end{subfigure}
        \hfill
        \begin{subfigure}[b]{0.16\textwidth}
            \includegraphics[width=\textwidth]{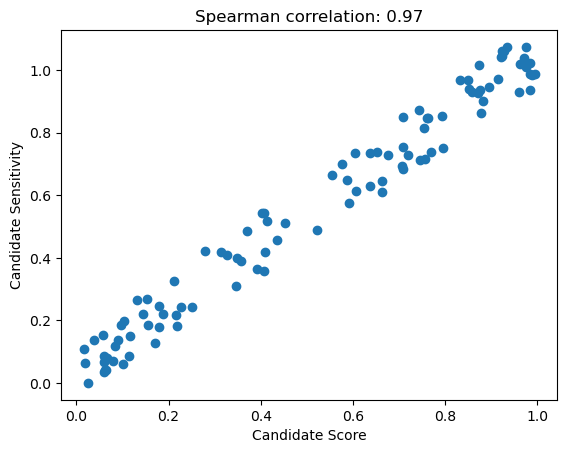}
        \label{fig:corr_linear_6}
        \end{subfigure}
        \hfill
    \end{subfigure}
    \begin{subfigure}[b]{\textwidth}
        \begin{subfigure}[b]{0.16\textwidth}
            \includegraphics[width=\textwidth]{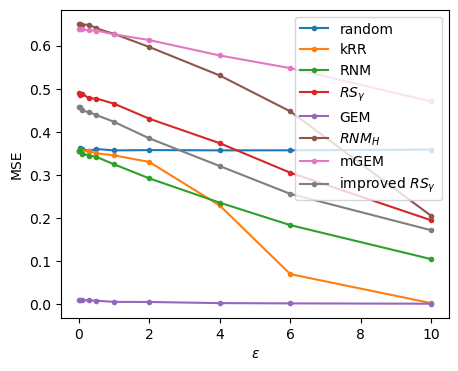}
        \label{fig:select_lin_0}
        \end{subfigure}
        \hfill
        \begin{subfigure}[b]{0.16\textwidth}
            \includegraphics[width=\textwidth]{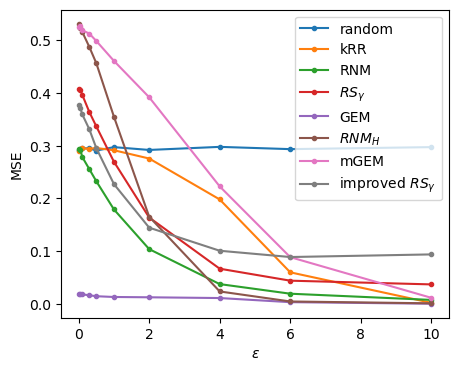}
        \label{fig:select_lin_1}
        \end{subfigure}
        \hfill
        \begin{subfigure}[b]{0.16\textwidth}
            \includegraphics[width=\textwidth]{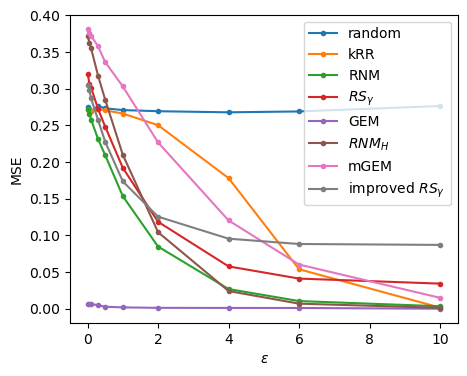}
        \label{fig:select_lin_2}
        \end{subfigure}
        \hfill
        \begin{subfigure}[b]{0.16\textwidth}
            \includegraphics[width=\textwidth]{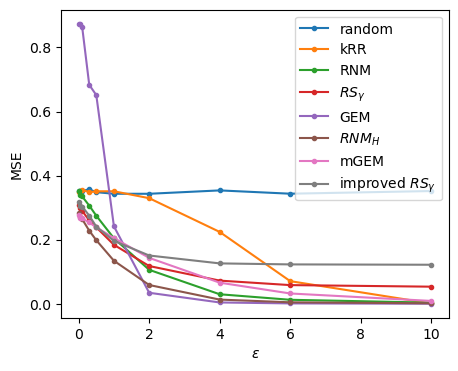}
        \label{fig:select_lin_4}
        \end{subfigure}
        \hfill
        \begin{subfigure}[b]{0.16\textwidth}
            \includegraphics[width=\textwidth]{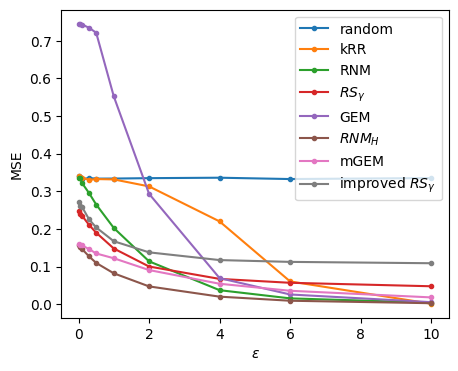}
        \label{fig:select_lin_5}
        \end{subfigure}
        \hfill
        \begin{subfigure}[b]{0.16\textwidth}
            \includegraphics[width=\textwidth]{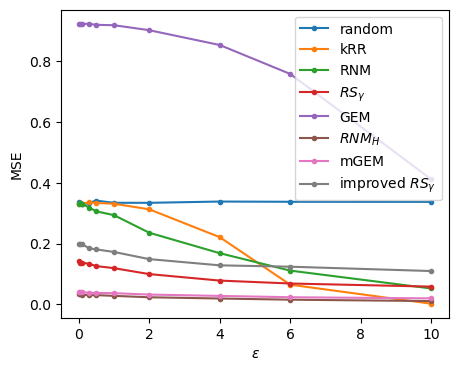}
        \label{fig:select_lin_6}
        \end{subfigure}
    \end{subfigure}
    \caption{Performance plots for increasing Spearman correlation in synthetic datasets with a linear relationship between scores and sensitivities. }
    \label{fig:increasing_corr}
\end{figure*}

We study three groups of scenarios. Our first set (Figure~\ref{fig:scenarios_bimodal_alt}) is the simple bimodal case discussed in the introduction. Our second set (Figure~\ref{fig:second_scenarios}) discusses scenarios similar to that of Figure~\ref{pictureofGEMscores}, confirming that the behavior in these settings is as expected. In the third set (Figure~\ref{fig:increasing_corr}) we look at settings where the sensitivities do not strictly increase or decrease with the scores. These experiments show that our heuristic remains effective in these more complex settings. 

In all experiments the number of candidates is set to $100$. The following hyper-parameters are used, unless specified otherwise: for $\rsy$, $\gamma = 0.05$; for GEM and mGEM, $\beta = 0.05$.

\paragraph{Scenarios 1, 2, 3: Bimodal scores.}
First, we consider simple scenarios with a bimodal split in the candidate scores and candidate-wise sensitivities: the high scores are set to $1$, the low to $-1$, and the sensitivities are either $1$ or $1.8$. There is no variability in the scores and sensitivities in this case. We look at positive, negative, and no correlation, respectively.

As expected based on the intuition we built earlier, scenario 1 has positive correlation so $\rsy$ and mGEM perform the best, followed by RNM, followed by GEM. The ordering is flipped in scenario 2 where we have negative correlation. We see the most dramatic deviation between the mechanisms for small $\epsilon$. 

In fact, as we see in both Figures \ref{fig:scenario_1}, and \ref{fig:scenario_2}, when $\epsilon$ is small, selecting the wrong heterogeneous noise mechanism can result in candidates from the lower scoring region being chosen most of the time and performance that is worse than random selection. For RS$_{\gamma}$, which behaves like RNMH, this is consistent with the discussion around eqn~\eqref{RNMHbehaviour}. For GEM and mGEM, this is consistent with the discussion in Section~\ref{syntheticdata}.

Scenario 3, where there is no correlation, behaves similarly to what we would expect if there was no heterogeneity in the candidate-wise sensitivities. The three mechanisms that utilise heterogeneity, $\rsy$, GEM, and mGEM, all perform slightly worse than RNM. This is likely due to not optimally utilising their privacy budget in this setting. Although there is no negative correlation, GEM outperforms $\rsy$ and mGEM, likely because it can utilise the fact that the optimal candidate has low sensitivity.
Appendix \ref{appendix:more_synthetic_experiments} shows further results for the bimodal setting with variable proportions of high and low scores.

\paragraph{Scenarios 4, 5, 6: More realistic scores} The next three scenarios simulate more realistic data, both by having more natural relationships between scores and sensitivities, and by allowing the scores to vary between trials. The particulars of each distribution are as follows. We will abuse notation and let $a$ denote both the candidate and the index of the candidate. In Scenario 4, we first sample 100 (the number of candidates) samples from $\mathcal{N}_{[0.01, 0.7]}(0.5; 1)$ and order these in increasing order $\sigma_1\le\sigma_2\le\cdots\le\sigma_{100}$ then for each candidate $a$, set $P_a = \mathcal{N}(\log(a), \sigma_a^2)$. In
Scenario 5, we set $P_a = \mathcal{N}(0.1a; 2.3 - 0.02a)$. Finally,
Scenario 6 first samples $\mu_a \sim U_{[0,1]}$ and $\sigma_a\sim\mathcal{N}_{[0.01, 0.7]}(0.5; 1)$ for each candidate then sets $P_a = \mathcal{N}(\mu_a, \sigma_a)$. Again, we observe (Figure \ref{fig:second_scenarios}) that, as expected, GEM does not perform well in the presence of positive correlation, but does in the presence of negative or absent correlation, and vice-versa for the mGEM and $\rsy$.

\paragraph{Scenarios with increasing correlation} In our final set of synthetic experiments, we explore how the gradual increase of correlation between scores and sensitivities affects mechanism performance. In this case, the scenarios are each indexed by a value $t \in [-5, -.8, -0.3, 0, 0.3, .8, 5]$. For each scenario (value of $t$), and each candidate, we sample $x_a \sim \mathcal{N}(0, 1)$, and $z_a \sim \mathcal{N}(0, 1)$ then define 
$P_a = \{(t * x + z)/5\} - \min_a\{(t * x + z)/5\}$. This results in distributions of candidate scores and sensitivities whose (Spearman) correlation goes from negative (left) to positive (right) in Figure \ref{fig:increasing_corr}. The bottom row of the figure shows the relative performance of the mechanisms in each scenario. As hypothesised, mGEM and $\rsy$ 
benefit most from strong positive correlation, and GEM from negative. The benefit of the best performing heterogeneous mechanism relative to $\rnm$ increases with the correlation strength.

\section{Choosing a Mechanism Adaptively}

We have thus far understood the correlation between scores and their sensitivities to be a useful heuristic for determining relative mechanism performance.
We will now explore whether this heuristic can be applied automatically, to (privately) infer the most appropriate selection mechanism to use at any given time. This allows us to combine GEM and mGEM into one mechanism that is designed to work reasonably well in all settings.

Given the candidate scores and candidate-wise sensitivities, we can leverage the correlation heuristic as follows. Intuitively, we privately report whether the scores are positively correlated with the sensitivities. If this is the case, mGEM completes the selection; otherwise, GEM completes the selection. The full algorithm, to which we refer as Combined GEM, is described in Algorithm~\ref{alg:combined_gem}. The private estimation of the indicator variable $\mathbf{1}_{[0,1]}(\text{spearman}(\mathbf{q}_{1:A}, \boldsymbol{\Delta}_{1:A}))$ is performed with randomised response ($kRR$ with $k=2$) \citep{Warner:1965} on the binary result of comparing the Spearman correlation to 0. We denote randomised response with a specific value of $\epsilon$ with $2RR_{\epsilon}$.

\begin{algorithm}

\caption{Combined GEM}\label{alg:combined_gem}
\begin{algorithmic}[1] 
\State \textbf{Inputs:} Privacy budgets $\epsilon_c$ and $\epsilon_g$, for correlation estimation and selection, respectively; access to $q(\cdot, D)$.

\If{$2RR_{\epsilon_c}(\text{spearman}(\mathbf{q}_{1:A}, \boldsymbol{\Delta}_{1:A}) \geq 0) = 1 $}
    \State Run mGEM with $\epsilon_g$
\Else
    \State Run GEM with $\epsilon_g$
\EndIf

\end{algorithmic}
\end{algorithm}

Spending some of the privacy budget $\epsilon_c$ on the mechanism choice leaves a budget of $\epsilon_g = \epsilon-\epsilon_c$ for the selection task. This means combined GEM trades off improved mechanism choice for performance in the selection mechanism itself. Thus, the benefit of this method will be mostly visible in datasets where the correlation can not be guessed in other ways. For example, in practice, it may be possible to infer the best choice of algorithm based on domain knowledge alone, without having to explicitly compute (and use extra budget for) the mechanism choice.

\begin{figure}
        \centering
        \captionsetup{justification=centering}
    \begin{subfigure}[b]{0.5\textwidth}
            \begin{subfigure}[b]{0.49\textwidth} 
                \includegraphics[width=.9\textwidth]
                {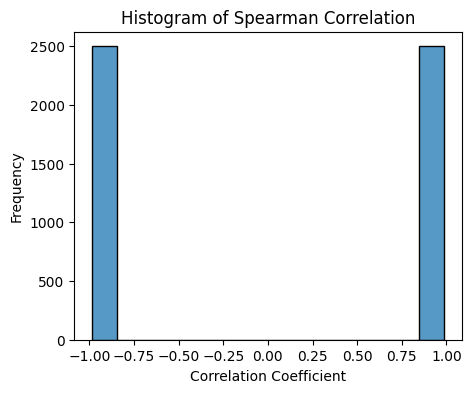}
                \caption{Scenario 7: a maximally polarised dataset}
                \label{fig:scenario7}
            \end{subfigure}
            \hfill
            \begin{subfigure}[b]{0.49\textwidth} 
                \includegraphics[width=.9\textwidth]{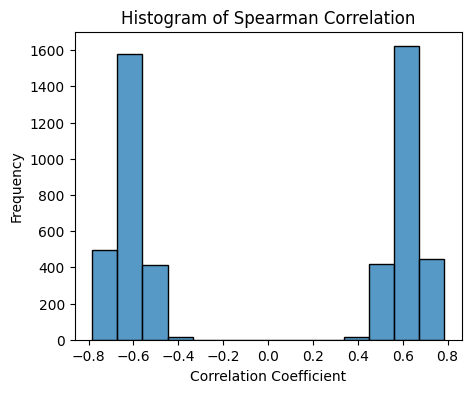}
                \caption{Scenario 8: a less polarised, noisier dataset}
                \label{fig:scenario8}
            \end{subfigure}

    \end{subfigure}

    \begin{subfigure}[b]{0.5\textwidth}
            
        \begin{subfigure}[b]{0.49\textwidth} 
                \includegraphics[width=.9\textwidth]
                {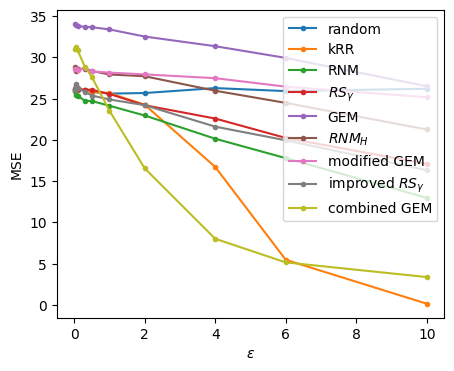}
                \caption{Performance of DP selection mechanisms vs privately adapting to data, Scenario 7}
                \label{fig:scenario7_selection}
            \end{subfigure}
            \hfill
            \begin{subfigure}[b]{0.49\textwidth} 
                \includegraphics[width=.9\textwidth]{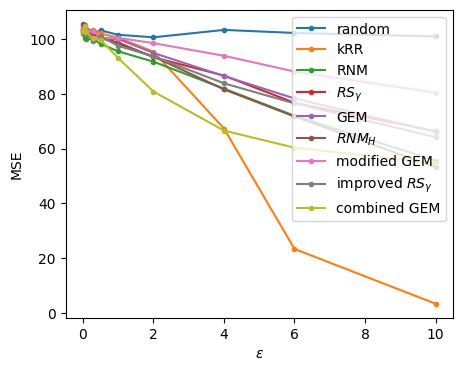}
                \caption{Performance of DP selection mechanisms vs privately adapting to data, Scenario 8}
                \label{fig:scenario8_selection}
            \end{subfigure}

    \end{subfigure}

    \caption{Evaluating the effect of privately adapting to the data versus using a single mechanism on the entire dataset.}
    \label{fig:combined_corrs}
\end{figure}

In Figure \ref{fig:combined_corrs}, we show this effect in two appropriate scenarios. In Scenario 7 shown in Figure \ref{fig:scenario7} we consider a case where half the datasets have strong positive correlation and half have strong negative negative correlation. In Scenario 8, Figure \ref{fig:scenario8}, shows a less polarised and noisier data set (data generation described in Appendix \ref{app:combined_data}). With $\epsilon_c=0.6\epsilon$, we observe that in scenario 7 choosing the mechanism adaptively based on the correlation heuristic can lead to significant performance improvements. As the correlations decrease in strength and the difference between GEM and mGEM decreases, spending budget on mechanism choice
may be less productive.\\
In the next section, we experiment with real world data for personalised recommendations. In these experiments we see that for most users, the scores and sensitivities are positively correlated. 
Consequently, in these tasks, it may be more beneficial to use domain expertise, or privately learn the average correlation across users, to select a single mechanism, rather than to spend budget on mechanism choice individually for each user. See Appendix \ref{appendix:amazon_books_demo_combined_gem}.

\section{Real-world data}
So far, we have investigated the behavior of private selection mechanisms in theory and on synthetic data. Next, we test our heuristic on real-world recommendation tasks. In several real-world datasets, we find correlation patterns which can be used to guide the best choice of private selection mechanism. 

Specifically, we find that scores predicted by SANSA \citep{Spivsak2023Scalable}, a state-of-the-art recommendation model, exhibit positive correlation with their associated sensitivities for a sizeable proportion of users in three real interaction datasets. This behaviour helps us to correctly anticipate the performance order and gap between the different DP selection mechanisms. In all three scenarios we find that our proposed mechanism mGEM is the best performing DP selection mechanism for small $\epsilon$. As expected, GEM can even perform worse than the random selection mechanism when $\epsilon$ is small.

\paragraph{Datasets} We experiment with three commonly used benchmark datasets: Netflix Prize data \footnote{\href{https://www.kaggle.com/datasets/netflix-inc/netflix-prize-data}{https://www.kaggle.com/datasets/netflix-inc/netflix-prize-data}}, MovieLens 20M \citep{movielens} \footnote{\href{https://www.kaggle.com/datasets/grouplens/movielens-20m-dataset}{https://www.kaggle.com/datasets/grouplens/movielens-20m-dataset}}, and the Amazon Books interactions dataset \footnote{\href{https://github.com/kuandeng/LightGCN/tree/master/Data/amazon-book}{https://github.com/kuandeng/LightGCN/tree/master/Data/amazon-book}}.
All datasets contain user IDs, item IDs (movies, books), and a binary interaction flag whenever a user interacted with an item. Any non-binary scores have been binarised according to \citet{Spivsak2023Scalable}.

\paragraph{Predicting scores with SANSA} SANSA, proposed by \citet{Spivsak2023Scalable}, is a scalable modification of a SOTA  shallow autoencoder architecture for collaborative filtering \citep{Steck2019Embarassingly}. It's attractive for its simplicity and good recommendations performance on naturally sparse datasets. Based on existing user-item interactions, $X$, for each user, the model predicts scores of unseen items by computing $\hat{\mathbf{q}} = \mathbf{u}^T B$, where $B$ is obtained by post-processing $A^{-1}=\left(X^T X+\lambda I\right)^{-1}$. SANSA proposes a sparse approximation to the matrix inversion.

\paragraph{Task and privacy model} Our goal is to understand the behaviour of different private selection mechanisms in a real-life task of recommending one item per user. To that effect, we replicate \citet{Spivsak2023Scalable} to (privately) predict item scores; then, for every user, we privately recommend the item with the highest predicted score.

We assume a privacy model in which all user-item engagements data should be DP-protected. Thus, private selection should be done in addition to private model training. In practice, training could be done using a DP algorithm, such as DP federated averaging \citep{abadi2016deep,mcmahan2017communication}, or by training on public or opt-in data. Since we are only interested in the performance of the selection mechanisms, we train SANSA by assuming a subset of each of the benchmarks data is public, and we use this as our training data. We show our results on the holdout set, which consists of tens of thousands previously unseen users\footnote{We use the test splits of \citet{Spivsak2023Scalable}.}. The item scores, which contain information about historical user-item engagements, should be privacy-protected with respect to a central server, which will observe all recommended items precisely but possibly in aggregate (i.e., the same privacy model as motivated by \citet{laro2023randomized}). Thus, we apply private selection algorithms on the set of candidate items and their SANSA scores. 

\paragraph{Experiment details} We replicate \citet{Spivsak2023Scalable}'s pre-processing to exclude any items with too few interactions. We use their test set only for the selection task.\footnote{On the Movielens dataset, our SANSA model has a recall of approx 0.36 vs 0.38 reported by \citet{Spivsak2023Scalable}, perhaps due to our usage of the SANSA library. This difference isn't relevant as our focus is the private selection task.} As before, we compare mechanisms on varying levels of epsilon, from 0.01 to 16. Sensitivities are computed as the maximum difference between the $1st$ and $99th$ percentiles of the predicted scores across all test users. The sensitivity of any candidate whose score is constant across all users is set to $10^{-6}$. Further, to mimic a realistic recommendation scenario, we limit the input set of candidate items per user to the top 500 for each user.\footnote{In practice, one would consume some privacy budget to do this. We do not include this in our privacy accounting since we are only interested in the relative performance of the selection mechanisms.}
Selection hyper-parameters are as follows: $\gamma = 0.008$ ($\rsy$)\footnote{Except for Netflix, where $\gamma = 0.0008$.}; $\beta = 0.05$ (GEM and mGEM).
For evaluation, we continue to use the MSE between the selected scores and the optimal scores.

\paragraph{Observed correlations and selection results} 
This recommendation task directly showcases the practicality of our heuristic. Each real datasets exhibits mostly positive correlation between scores and sensitivities. As expected, this results in mGEM being the best performing mechanism on average. 

\begin{figure*}[htbp]
    \centering
    \begin{subfigure}[b]{0.25\textwidth}
        \includegraphics[width=\textwidth]{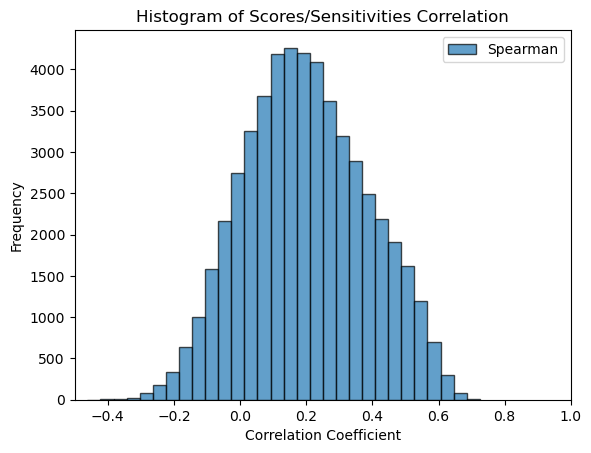}
    \caption{Amazon Books correlations}
    \label{fig:books_predicted_correlations}
    \end{subfigure}
    \hfill
    \begin{subfigure}[b]{0.25\textwidth}
        \includegraphics[width=\textwidth]{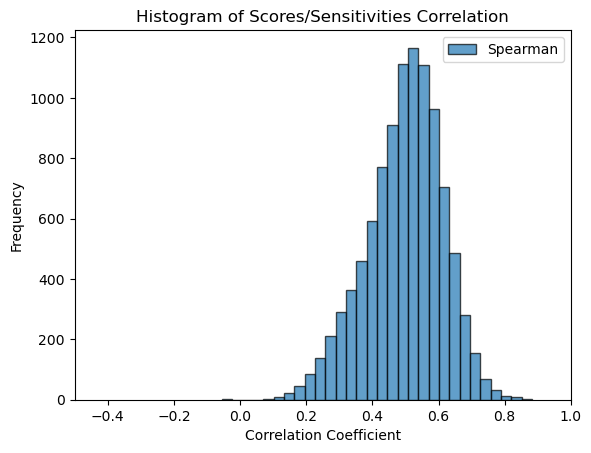}
    \caption{MovieLens20 correlations}
    \label{fig:movielens_correlations}
    \end{subfigure}
    \hfill
    \begin{subfigure}[b]{0.25\textwidth}
        \includegraphics[width=\textwidth]{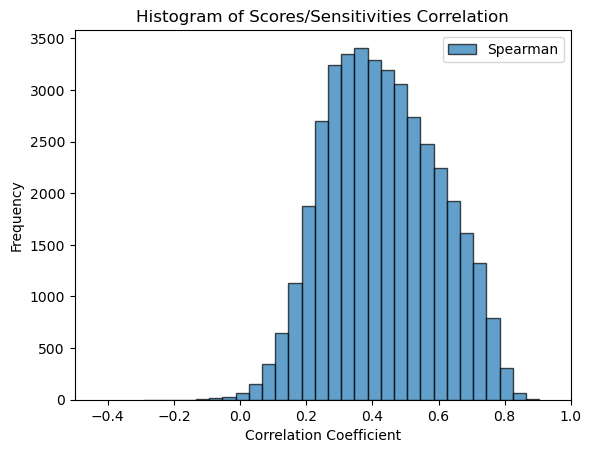}
    \caption{Netflix Prize correlations}
    \label{fig:netflix_predicted_correlations}
    \end{subfigure}
     \hfill

    \begin{subfigure}[b]{0.25\textwidth}
       \includegraphics[width=\textwidth]{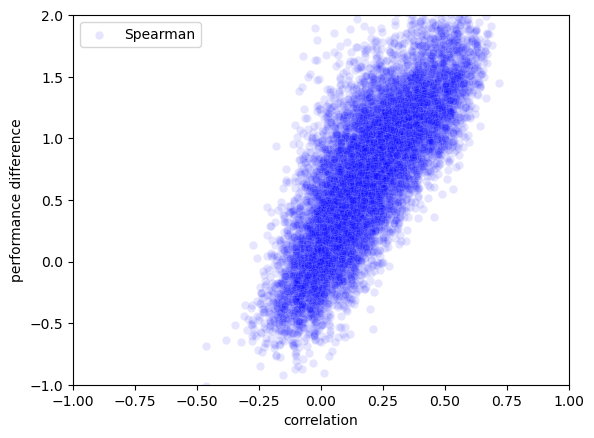}
        \caption{Amazon Books mGEM/GEM vs correlation}
        \label{fig:books_scatterplot}
    \end{subfigure}
    \hfill
     \begin{subfigure}[b]{0.25\textwidth}
       \includegraphics[width=\textwidth]{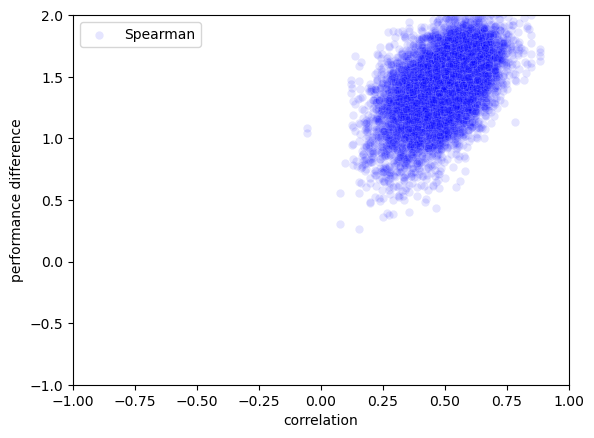}
        \caption{Movielens20 mGEM/GEM vs correlation}
        \label{fig:movielens_scatterplot}
    \end{subfigure}
    \hfill
    \begin{subfigure}[b]{0.25\textwidth}
       \includegraphics[width=\textwidth]{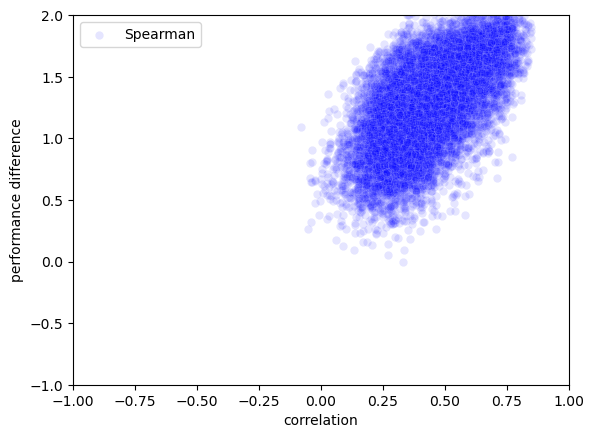}
        \caption{Netflix mGEM/GEM vs correlation}
        \label{fig:netflix_scatterplot}
    \end{subfigure}

    \begin{subfigure}[b]{0.25\textwidth}
       \includegraphics[width=\textwidth]{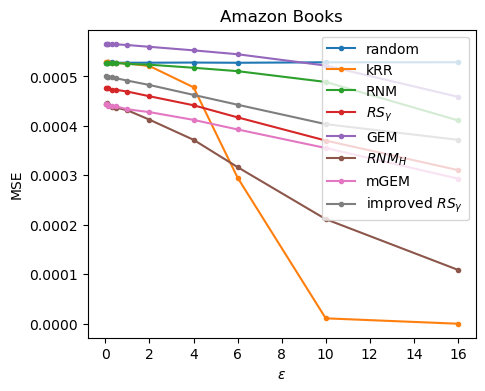}
        \caption{Amazon Books selection}
        \label{fig:books_selection}
    \end{subfigure}
    \hfill
    \begin{subfigure}[b]{0.25\textwidth}
       \includegraphics[width=\textwidth]{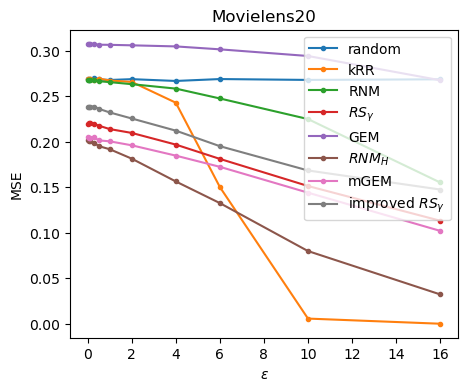}
        \caption{Movielens20 selection}
        \label{fig:movielens_selection}
    \end{subfigure}
    \hfill
    \begin{subfigure}[b]{0.25\textwidth}
       \includegraphics[width=\textwidth]{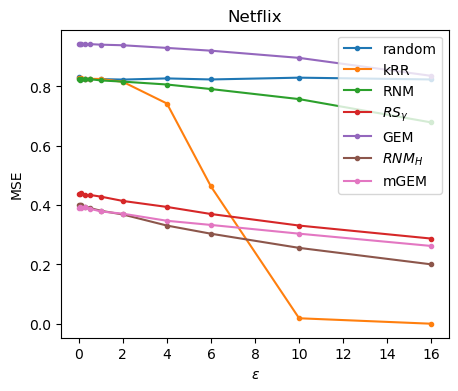}
        \caption{Netflix selection}
        \label{fig:netflix_selection}
    \end{subfigure}
\caption{Comparison of private selection mechanisms in a recommendation task with real datasets. The top row shows the distribution of Spearman correlation coefficients (between each users' predicted item scores and those items' sensitivities) across all users. The second row plots the performance difference between GEM and mGEM, per user, as a function of their calculated correlation coefficient. The performance gap between two mechanisms is measured as the log ratio between the mean selected scores of the two mechanisms over 50 trials. For clarity of the scatter plot, only 10,000 random test users are shown. The bottom row shows the MSE (of the selected score relative to the highest score) measured when running different selection mechanisms on the predicted scores for the entire test set.}\label{realdata}
\end{figure*}

We investigate the correlation in two ways. Figures 
\ref{fig:books_predicted_correlations}, \ref{fig:movielens_correlations}, and \ref{fig:netflix_predicted_correlations} shows the distribution of the Spearman's rank correlation coefficient over all users. We observe some positive correlation in all three datasets (Netflix has most). Figures \ref{fig:books_scatterplot}, \ref{fig:movielens_scatterplot}, and \ref{fig:netflix_scatterplot} show the relative performance of mGEM and GEM plotted against Spearman's rank correlation coefficient. Each dot corresponds to a single user, 10,000 random test users are shown. We observe that, for each dataset, the performance difference between the algorithms grows as the correlation coefficient grows, as expected. Amazon Books is the only dataset with any users showing negative correlation between their possible items' scores and sensitivities; here, we see that the trend also holds when the domain is negative: the more negative the correlation the better GEM is compared to mGEM.

Finally, Figures \ref{fig:books_selection}, \ref{fig:movielens_selection}, and \ref{fig:netflix_selection} show the aggregate performance of the various private selection mechanisms. 
Both the order and the gap between the mechanisms is as expected: more users with stronger positive correlation benefits mGEM and $\rsy$.
GEM performs worse than RNM in all three datasets.

\section{Outlook: Benefits of Private Selection under Distribution Shift}
\label{s:distr_shift}
Building on these insights, we now explore a broader application of private selection mechanisms – online learning – where we hypothesise that ensuring differential privacy can simultaneously facilitate exploration. Specifically, we examine how the mechanisms studied thus far can address challenges arising in sequential decision-making tasks, under conditions of distribution shift. In fact, DP selection mechanisms often appear as subroutines as part of online learning problems where one sequentially selects candidates, then updates the candidate score function after receiving some feedback about the prior selection. In these online problems where one is training the score function over time,  there is often a trade-off between exploration (choosing actions to improve model accuracy) and exploitation (choosing actions to improve realised rewards). This trade-off can be especially pertinent under cold-start and distribution shift settings. Furthermore, algorithms that maintain distinct confidence intervals per candidate, such as Thompson Sampling and Upper Confidence Bound (UCB), often outperform algorithms that assume homogeneous confidence intervals or ignore them altogether, such as epsilon-greedy or softmax \citep{bietti2021contextual}. In this exploratory section, we consider two research questions:

\begin{enumerate}
\item In the presence of distribution shift, might private selection mechanisms incur less cumulative regret compared to non-private mechanisms?
\item Do our findings from the previous sections (namely, that mGEM outperforms all other private algorithms under the positive correlation setting) hold in the online learning setting with distribution shift?
\end{enumerate}
\begin{figure*}[htbp]
        \centering
        \begin{subfigure}[b]{0.24\textwidth}
            \centering
            \includegraphics[width=\textwidth]{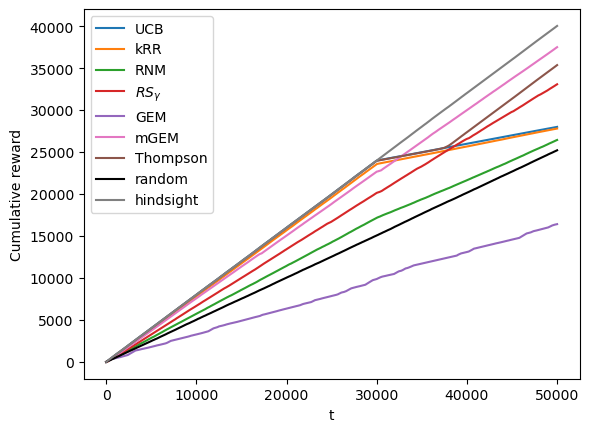}
            \caption{Cumulative reward}
            \label{fig:synth_online_reward}
        \end{subfigure}
        \hfill
        \begin{subfigure}[b]{.24\textwidth}  
            \centering 
            \includegraphics[width=\textwidth]{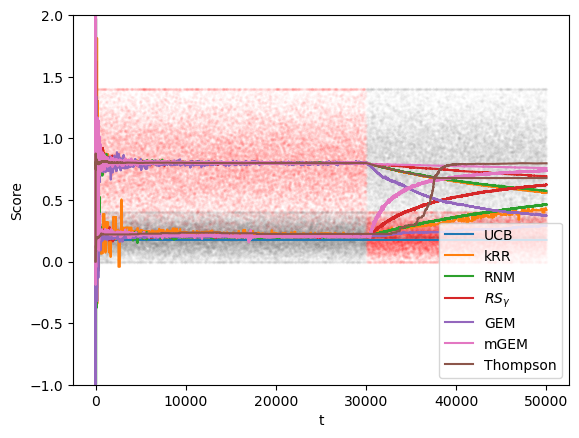}
            \caption{Estimated means}
            \label{fig:synth_online_means}
        \end{subfigure}
    \hfill
        \begin{subfigure}[b]{.24\textwidth}  
            \centering 
            \includegraphics[width=\textwidth]{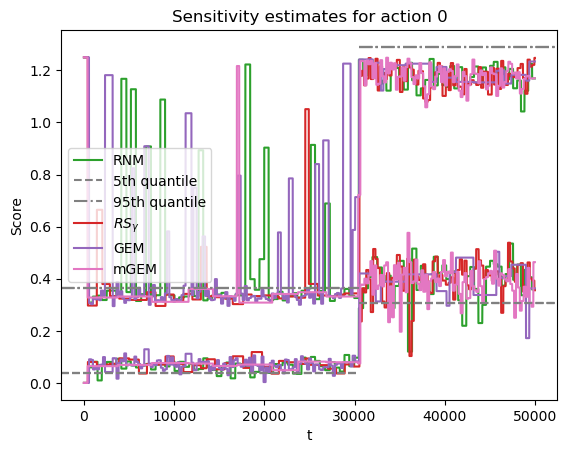}
            \caption{Estimated quantiles}
            \label{fig:q_a0}
        \end{subfigure}
    \hfill
        \begin{subfigure}[b]{.24\textwidth}  
            \centering 
            \includegraphics[width=\textwidth]{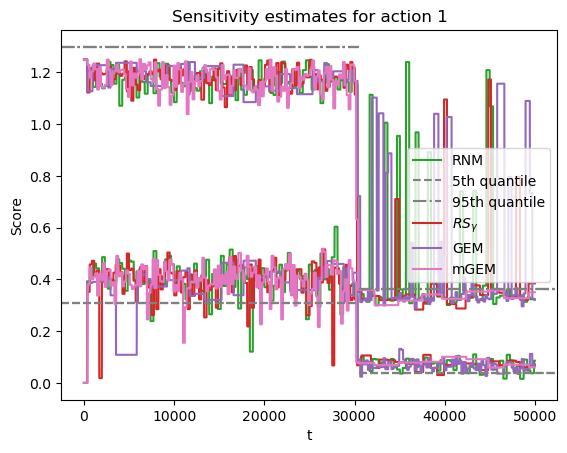}
            \caption{Estimated quantiles}
            \label{fig:q_a1}
        \end{subfigure}
    \caption{Comparison of non-private UCB and TS vs. private kRR, RNM, $\rsy$ ($\gamma=.05$), GEM ($\beta=0.05$), and mGEM ($\beta=0.05$) as selection mechanisms in an online learning, multi-armed bandit setting. Distribution shift happens at $t_{shift}=30,000$ and the horizon is $T=50,000$. (a) plots cumulative realised reward per algorithm over time. Note that GEM performs poorly due to the positive correlation between scores and sensitivities. (b) plots the estimated means (solid lines) of the different algorithms transposed over the private scores (points) for action \textcolor{red}{$0$} and \textcolor{gray}{$1$}. (c) and (d) plot the privately estimated sensitivity quantiles of each action, demonstrating that they are approximately accurate (close to the dashed lines).
    }
    \label{fig:shifted_bandits}
\end{figure*}
There are two important differences between non-private algorithms like Thompson Sampling and UCB and the heterogeneous private algorithms discussed herein. First, there is no obvious relationship between per-action sensitivities and expected reward posterior distributions, meaning that the realised performances of the private and non-private variants may be arbitrarily different. Second, ensuring differential privacy can be thought of as enforcing a minimal amount of exploration. In this sense, it is obvious that private selection will incur an efficiency cost in non-adversarial settings. However, in practical settings, where distribution shift is common, we hypothesize that maintaining some fixed amount of exploration per action will actually lead to efficiency gains. Therefore, we perform a bandit simulation (without context) to investigate the relative performance under distribution shift of a) non-private UCB and Thompson Sampling, b) private selection with homogeneous noise (RNM, RR), and c) private selection with heterogeneous noise (GEM, mGEM, and RS$_\gamma$). We hypothesize that the private selection algorithms might actually achieve lower cumulative regret than the non-private baseline. We provide our detailed experiment methodology in \ref{appendix:distr_shift}. Our problem has two possible actions which we label 0 and 1. While we use bandit learners in the present simulations for simplicity, the ideas could be extended to the contextual bandit setting by pairing a private contextual bandit learner under a joint \citep{shariff2018differentially} or local \citep{zheng2020locally,garcelon2021Local,han2021Generalized}) privacy setting with one of the private selection mechanisms discussed herein. This would ensure that actions can be observed precisely by a third party (e.g. the server) without the loss of user privacy.

\paragraph{Results}
In Figure \ref{fig:synth_online_reward} we plot the cumulative reward obtained by using each policy. As hypothesised, we find that all DP variants outperform the non-private UCB baseline after the distribution shift occurs. On inspection, we find that the UCB bandit keeps picking action 1 for all time steps in the simulation even once it is the sub-optimal action. This is not unexpected since UCB regret bounds are proven only in the non-adversarial case (no distribution shift). We note that both before and after the distribution shift, the scores and candidate sensitivities have positive correlation. The mGEM mechanism is found to outperform all others, including original GEM, $\rsy$, kRR, and RNM. Heterogeneous noise proves to be beneficial from the start, but in addition it also appears to speed up the adaptation to the shift in distribution at $t=500$. The ability to adapt faster to shifting distributions is again showcased in Figure \ref{fig:synth_online_means}, where the predicted mean scores together with the private scores of candidates $0$ and $1$ are plotted over time. For mGEM and $\rsy$, the predicted means adjust more rapidly and they target action $1$. These experiments provide evidence that a) private selection can reduce regret under distribution shift and b) mGEM can significantly outperform GEM and other private selection algorithms, depending on the dataset characteristics.

\section{Conclusion}

The theoretical and empirical results that we've presented suggest that correlation between the candidate scores and sensitivities is a useful heuristic for determining which DP selection mechanism will perform well for a given selection task. In particular, mGEM is a consistently good choice when the correlation is positive while GEM is a consistently good choice when the correlation is negative. Towards the goal of combining the best of both, we proposed combined GEM, which adapts to the data to choose the appropriate GEM version. We recommend this mechanism whenever the correlation can not be estimated by other means. Furthermore, we found that, in the real recommendation tasks we studied, predicted score and sensitivities were indeed positively correlated, and our proposed mechanism mGEM outperformed all other algorithms. 
We conjecture that the positive correlation setting may be common in real-world applications. Therefore, while the best algorithm will depend on the specifics of the data distribution at hand, we hypothesize there are many practical use cases in which mGEM would be a strong choice. Since each of GEM, mGEM and $\rsy$ can perform poorly in ome settings, when correlation is weak or unknown, RNM remains a safer option.

\newpage

\bibliographystyle{plainnat} 
\bibliography{references_editable}

\appendix

\subsection{Counterexample for RNM with Heterogeneous Noise}
\label{appendix:example}

We showed in Section~\ref{RNMHnotDP} that RNMH is not differentially private.  

In this section, we provide a counterexample to show RNMH with Laplacian noise instead of Exponential noise is not $\epsilon'$-differentially-private for any $\epsilon'<(k-1)\epsilon$, where $k$ is the number of candidates. Consider RNMH with Laplace noise, i.e. the noised scores are $\noisyscore=q_a + z_a $ with $z_a \sim \text{Lap}(\Delta_i/\epsilon)$ and $\epsilon>0$. Now in dataset $D_1$ we have $q_i=0$ for $i=1,...,k$ and in dataset $D_2$ we have $q_1=0$ but $q_i=1$ for $i=2,...,k$. Suppose candidate one has sensitivity $\Delta_1=0$ and candidates $2,...,k$ have candidate-wise sensitivities $\Delta_i=1$.

We now compute the probability of RNMH with Laplacian noise outputting candidate 1 on both datasets. For dataset $D_1$ we have that
\begin{align}
    Pr(M(D_1)=1)    &= Pr(z_i<0 \text{ for } i=2,...,k)\\
                    &= \prod_{i=2}^k Pr(z_i<0)\\
                    &= \left(\frac{1}{2}\right)^{k-1}
\end{align}
On the other hand we have that
\begin{align}
    Pr(M(D_2)=1)&= Pr( z_i<-1 \text{ for } i=2,...,k)\\
                &= \prod_{i=2}^k Pr(z_i<-1)\\
                &= \left(\frac{1}{2} \exp(-\epsilon)\right)^{k-1}
\end{align}
Thus, 
\begin{align}
    \frac{Pr(M(D_1)=1)}{Pr(M(D_2)=1)} &= \left(\frac{\frac{1}{2}}{\frac{1}{2}\exp(-\epsilon)}\right)^{k-1}\\
                    &=\exp((k-1)\epsilon),
\end{align}
which concludes the counterexample.

\subsection{Counterexample for $\rsy$ with Exponential Noise}
We know $\rsy$ is $\epsdp$ when the noising mechanism used for each candidate is DP. Given that RNM with Exponential noise is DP, a natural question is whether $\rsy$ is also $\epsdp$ when using Exponential noise instead of the Laplace distribution. Here we give a counterexample to show that $\rsy$ with Exponential noise is not DP. 
Consider datasets $D_1$ and $D_2$ such that in dataset $D_1$ we have $q_i=1$ for $i=1,...,k$ and in dataset $D_2$ we have $q_1=1$ but $q_i=0$ for $i=2,...,k$. Suppose candidate one has sensitivity $\Delta_1=0$ and candidates $2,...,k$ have candidate-wise sensitivities $\Delta_i=1$.

Then we have that under dataset $D_1$ the only way to output candidate 1 is to only sample that candidate so
$$Pr(M(D_1)=1)=\sum_{t=1}^{\infty} (1-\gamma)^t \left(\frac{1}{k}\right)^{t} = \frac{(1-\gamma)\frac{1}{k}}{1- (1-\gamma)\frac{1}{k}}.$$

On the other hand, for $D_2$ it outputs candidate 1 whenever candidate 1 is sampled and all other noise samples are smaller $1$. This probability is given by
\begin{align*}
    &Pr(M(D_2)=1) \\
    &=\sum_{t=1}^{\infty} (1-\gamma)^t \left( \sum_{s=0}^t Pr(\text{Bin}(t, 1-\frac{1}{k})=s)\left( Pr(Z \leq 1)\right)^s \right)\\
    &=\sum_{t=1}^{\infty} (1-\gamma)^t \left( \sum_{s=0}^t Pr(\text{Bin}(t, 1-\frac{1}{k})=s)\left( 1-e^{-\epsilon/6}\right)^s \right)\\
    &=\sum_{t=1}^{\infty} (1-\gamma)^t \left( \sum_{s=0}^t \binom{t}{s}\left(1-\frac{1}{k}\right)^s \frac{1}{k^{t-s}}\left( 1-e^{-\epsilon/6}\right)^s \right)\\
    &=\sum_{t=1}^{\infty} (1-\gamma)^t \frac{1}{k^t} \left( \sum_{s=0}^t \binom{t}{s}\left(k-1\right)^s \left( 1-e^{-\epsilon/6}\right)^s \right)\\
    &=\sum_{t=1}^{\infty} (1-\gamma)^t \frac{1}{k^t} \left( 1+ \left(k-1\right) \left( 1-e^{-\epsilon/6}\right) \right)^t\\
    &=\frac{(1-\gamma) \frac{1}{k} \left( 1+ \left(k-1\right) \left( 1-e^{-\epsilon/6}\right) \right)}{1-(1-\gamma) \frac{1}{k} \left( 1+ \left(k-1\right) \left( 1-e^{-\epsilon/6}\right) \right)}\\
\end{align*}
where $Z\sim \text{Exp}(\epsilon/6)$ in an attempt to match random stopping.

Taking the ratio of the two probabilities we get
\begin{align*}
    &\frac{Pr(M(D_2)=1)}{Pr(M(D_1)=1)} = \\
    &\frac{1+ \left(k-1\right) \left( 1-e^{-\epsilon/6}\right)}{1-(1-\gamma) \frac{1}{k} \left( 1+ \left(k-1\right) \left( 1-e^{-\frac{\epsilon}{6}}\right) \right)}  \left(1-(1-\gamma) \frac{1}{k}\right)\\.
\end{align*}
We note that for any fixed $\epsilon>0$ this ratio tends to infinity as $k\rightarrow \infty$, which concludes the counterexample.

\subsection{Improved version of Randomised Stopping}
\label{app:improved_rs}
For $\rsy$ the number of samples taken, here denoted by $K$, is sampled from the geometric distribution, i.e. $K \sim Geometric(\gamma)$, where $\gamma$ is the probability of a success. Instead, \cite{papernot2022hyperparameter} propose to sample $K$ from the truncated negative binomial distribution defined below. Corollary 3 in \cite{papernot2022hyperparameter} states that for base mechanisms, which are $\epsilon$-DP, randomized stopping with the truncated negative binomial distribution satisfies $(2+\eta)\epsilon$-DP. In all our experiments we set $\eta=0$. For $\eta=1$ the privacy bound of \cite{Liu_2019PrivateCandidates} is recovered.

\begin{definition}[Truncated Negative Binomial Distribution]
 Let $\gamma \in(0,1)$ and $\eta \in(-1, \infty)$. Define a distribution $\mathcal{D}_{\eta, \gamma}$ on $\mathbb{N}=\{1,2, \cdots\}$ as follows. If $\eta \neq 0$ and $K$ is drawn from $\mathcal{D}_{\eta, \gamma}$, then

$$
\forall k \in \mathbb{N} \quad \mathbb{P}[K=k]=\frac{(1-\gamma)^k}{\gamma^{-\eta}-1} \cdot \prod_{\ell=0}^{k-1}\left(\frac{\ell+\eta}{\ell+1}\right)
$$

and $\mathbb{E}[K]=\frac{\eta \cdot(1-\gamma)}{\gamma \cdot\left(1-\gamma^\eta\right)}$. If $K$ is drawn from $\mathcal{D}_{0, \gamma}$, then

$$
\mathbb{P}[K=k]=\frac{(1-\gamma)^k}{k \cdot \log (1 / \gamma)}
$$

and $\mathbb{E}[K]=\frac{1 / \gamma-1}{\log (1 / \gamma)}$.
\end{definition}

\subsection{Additional Results in the two Candidate Setting}
\label{appendix:two_candidates}

\paragraph{Hyper-parameter Ablations.} 

\begin{figure*}[htbp]
\centering
\includegraphics[width=.8\textwidth]{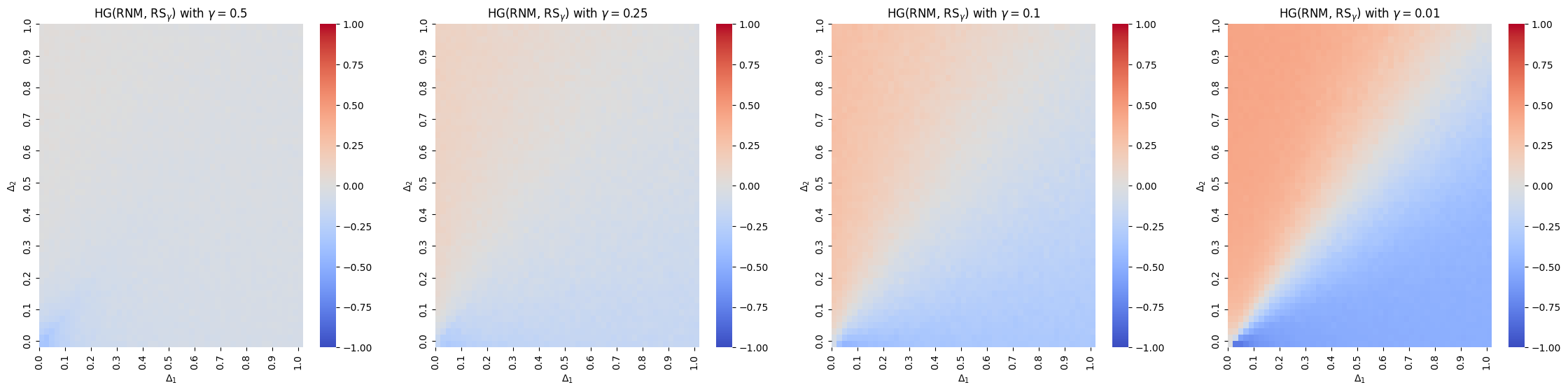}
\caption{Plotting $HG(\rnm, \rsy)$ for varying levels of $\gamma$ and at different sensitivities $\Delta_1$ on the x-axis and $\Delta_2$ on the y-axis. All values of $HG(\rnm,\rsy)$ are computed for $\epsilon=0.1$ and a scores $q_1=0, q_2=1$. As $\gamma$ decreases and candidates are sampled several times a stronger effect of heterogeneity and the correlation between scores and sensitivities on the performance is observed.}
\label{fig:hg_rsy_gammas}
\end{figure*}

As $\gamma$ controls the number of samples taken by $\rsy$, it has a strong effect on the performance. In Figure \ref{fig:hg_rsy_gammas} we performed an ablation to investigate the effect of $\gamma$ on the relative performance of $\rsy$ against $\rnm$. We observe that for large $\gamma$ as the expected number of samples is $2$, there are only small performance differences between $\rnm$ and $\rsy$. However, as $\gamma$ increases, the differences between the two mechanisms become more visible. If gamma is small relative to the number of candidates, i.e. when $1/\gamma>>|\actionspace|$, candidates are very likely to be sampled multiple times. This results the difference between RNM and $\rsy$ being pronounced. 

\begin{figure*}[htbp]
\centering
\includegraphics[width=\textwidth]{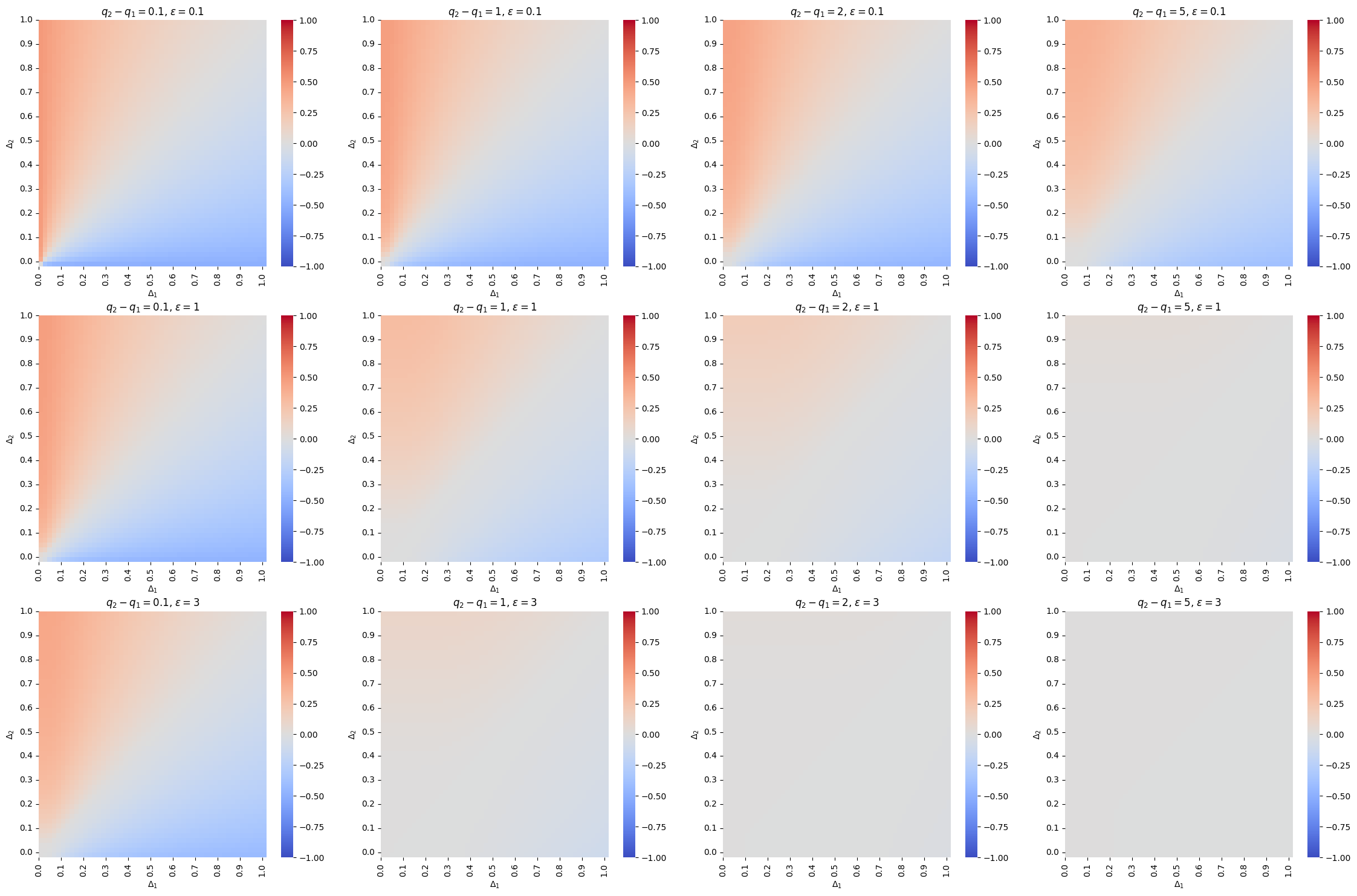}
\caption{Plotting $HG(\rnm,\rnmh)$ against $\Delta_1$ on the x-axis and $\Delta_2$ on the y-axis. Within each row the difference between the scores $q_2-q_1$ takes values $0.1,1,2,5$ from left to right and the value of $\epsilon$ stays the same. Across rows $\epsilon=0.1,1,3$ from top to bottom.}
\label{app:fig:Hg_RNMH}
\end{figure*}

\begin{figure*}[htbp]
\centering
\includegraphics[width=\textwidth]{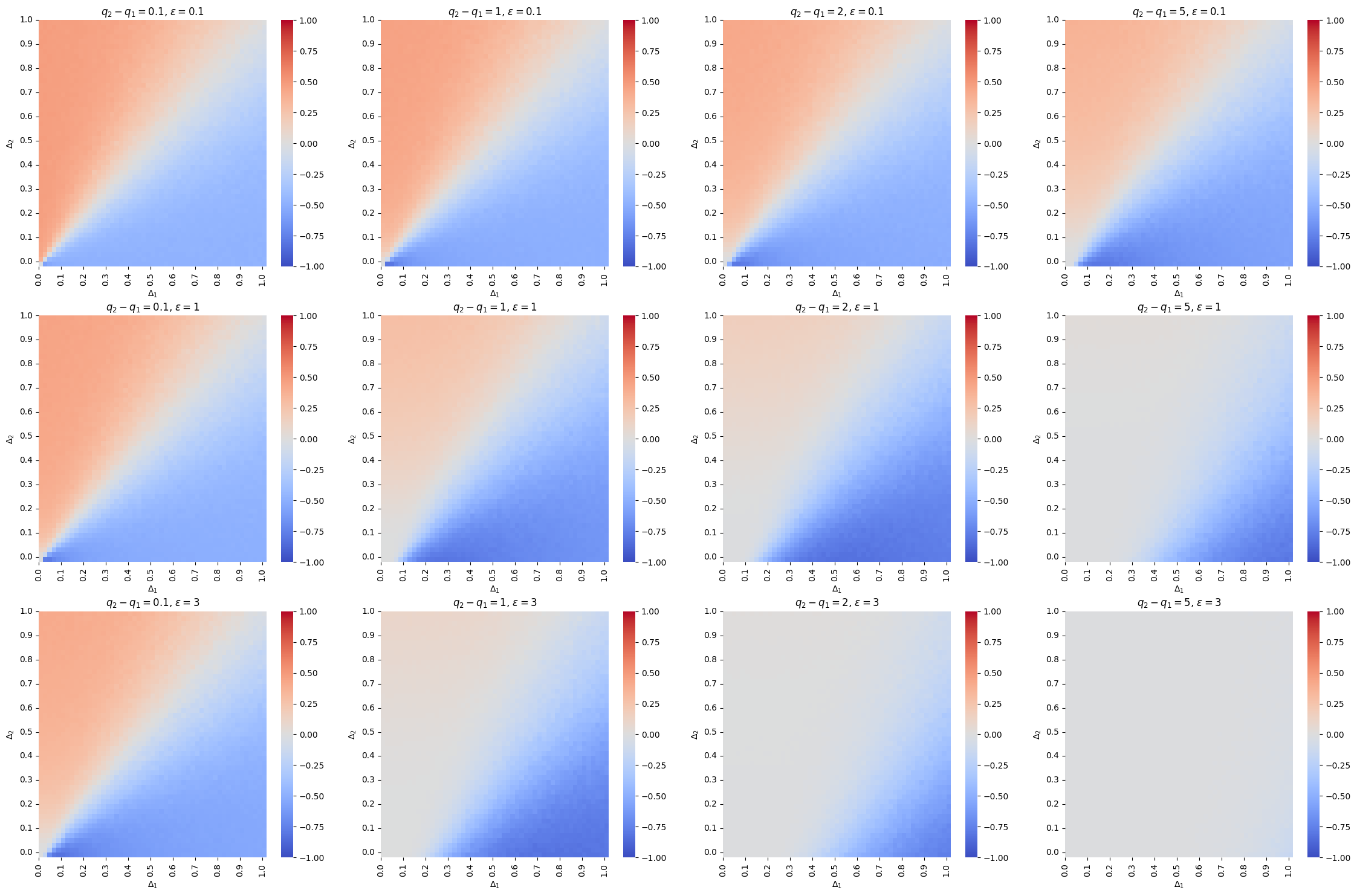}
\caption{Plotting $HG(\rnm,\rsy)$ with $\gamma=0.01$ plotted against $\Delta_1$ on the x-axis and $\Delta_2$ on the y-axis. Within each row the difference between the scores $q_2-q_1$ takes values $0.1,1,2,5$ from left to right and the value of $\epsilon$ stays the same. Across rows $\epsilon=0.1,1,3$ from top to bottom.}
\label{app:fig:Hg_RSy}
\end{figure*}

\paragraph{Ablations on $\epsilon$ and $q_2-q_1$.}
In Figures~\ref{app:fig:Hg_RNMH} and \ref{app:fig:Hg_RSy} we investigate the relative performance of $\rnmh$ and $\rsy$ against RNM as $\epsilon$ and the difference between the two scores are varied. 

In each figure, the blue regions correspond to regimes where RNM outperforms the comparison algorithm.


\subsection{Additional Synthetic Data Experiments}
\label{appendix:more_synthetic_experiments}

\begin{figure*}
    \centering
    \begin{subfigure}[b]{0.32\textwidth}
        \includegraphics[width=\textwidth, trim={0 0 0 0cm},clip]{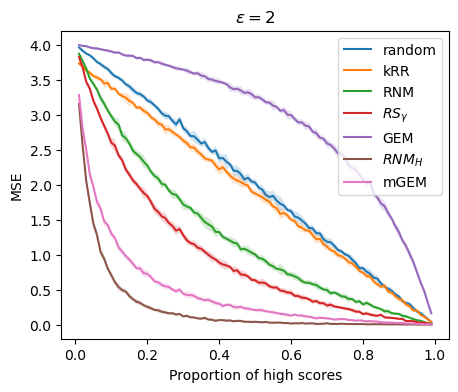}
    \caption{Scenario 1: high scores = 1, sensitivities = $1.8$; low scores = -1, sensitivities = 1.}
    \end{subfigure}
    \hfill
    \begin{subfigure}[b]{0.32\textwidth}
       \includegraphics[width=\textwidth, trim={0 0 0 0cm},clip]{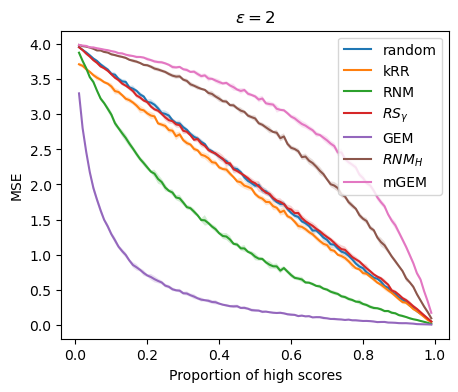}
       
        \caption{Scenario 2: same means, but the high score class now has small sensitivities.}
    \end{subfigure}
    \hfill
    \begin{subfigure}[b]{0.32\textwidth}
       \includegraphics[width=\textwidth]{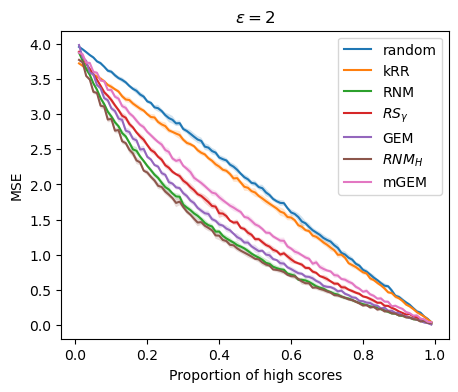}
        \caption{Scenario 3: same means, but with an equal share of large and small sensitivities.}
    \end{subfigure}
    
    \caption{Varying the proportion of high scores out of all scores for Scenario 1, 2 and 3. Positive correlation between the scores and sensitivities is beneficial to  mGEM; negative correlation is beneficial to GEM. The gap in performance is largest when there are a few high scores.}
    \label{fig:varying_prob_alt}
\end{figure*}

In Figure \ref{fig:varying_prob_alt} we set $\epsilon = 2$ and vary the proportion of high scores out of all scores in each of the three scenarios. This allows us to go beyond the equal split in the previous figures and observe how the gap in performance between the algorithms changes with the number of high scores. As expected, in the positive correlation case, RNMH, mGEM, and $\rsy$ perform better than the homogeneous alternatives or the standard GEM, for most proportions of high scores. The gap in performance is smallest at the extremes. In the case of negative correlation, the use of heterogeneous noise is only beneficial in GEM. The other heterogeneous algorithms perform worse than random across all non-trivial splits between high and low scores at this epsilon value. When there's no correlation and the best candidate has a sensitivity comparable to the rest, it's not beneficial to use heterogeneity and RNM suffices at all splits.

\subsection{Combined Gem Synthetic Dataset}
\label{app:combined_data}

\begin{figure}
    \centering
    \includegraphics[width=0.9\linewidth]{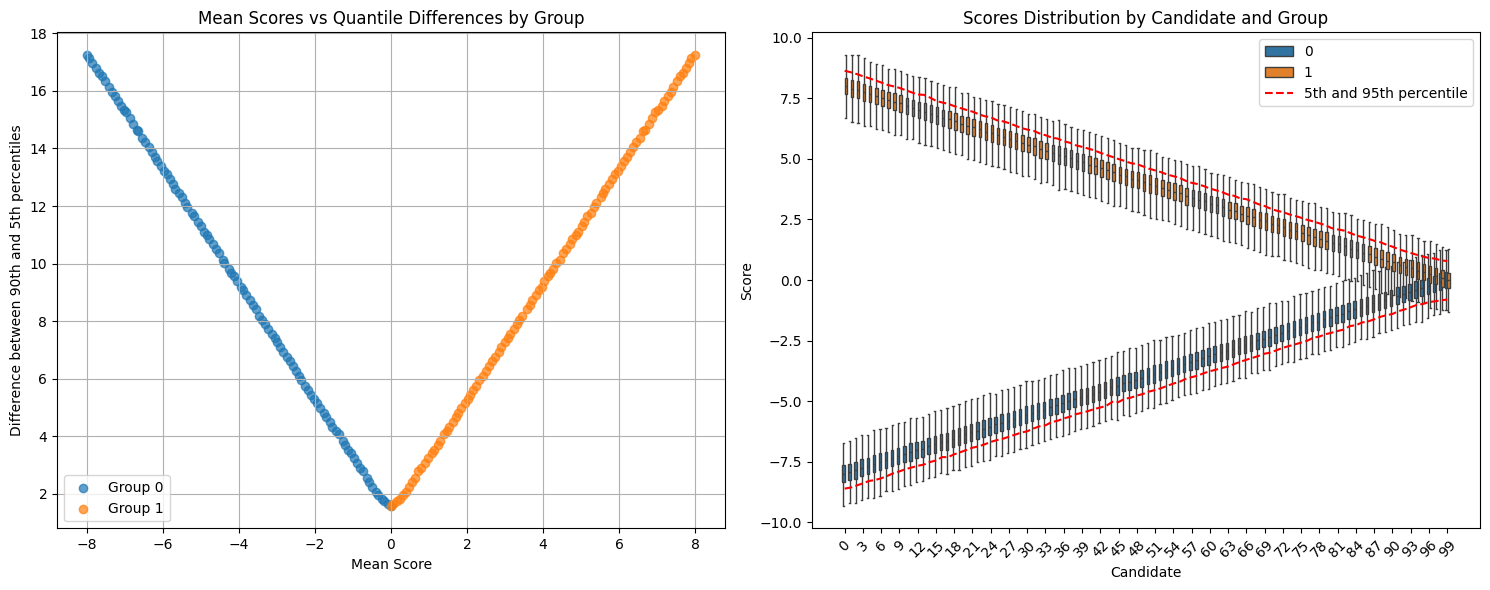}
    \caption{Strong polarization setting. Boxplots showing the within group distribution of the scores across candidates. Red lines indicate the 5th and 95th quantiles.}
    \label{fig:combined_strong}
\end{figure}

\begin{figure}
    \centering
    \includegraphics[width=0.9\linewidth]{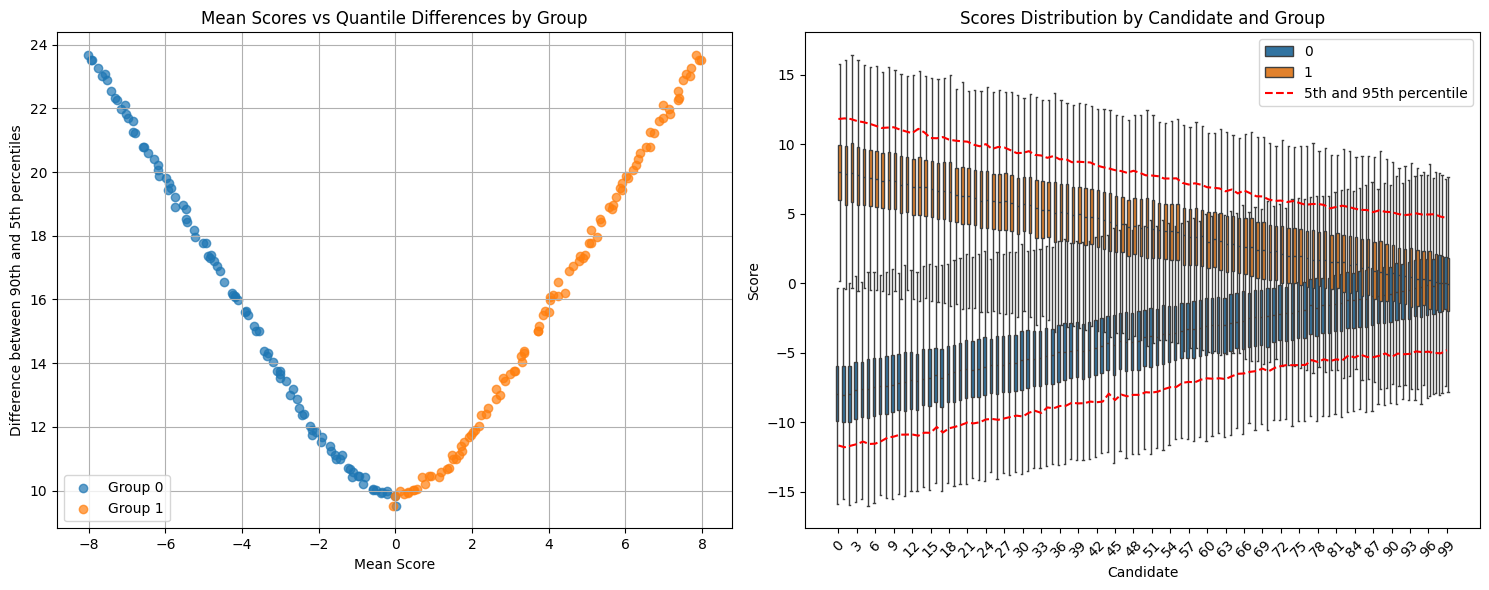}
    \caption{Weak Polarization setting. Boxplots showing the within group distribution of the scores across candidates. Red lines indicate the 5th and 95th quantiles.}
    \label{fig:combined_weak}
\end{figure}

The synthetic dataset used for the comparison of combined GEM with other selection mechanisms has been generated to contain users with strong positive as well as strong negative correlation. Figure~\ref{fig:combined_strong} and Figure~\ref{fig:combined_weak} show the sensitivities across mean scores by groups as well as the distribution of scores across candidates. The datasets were generated as follows: In our experiments we use a total of $N=5000$ users and $|\actionspace|=100$ candidates and users are split equally into two groups $0$ and $1$, denoted by $g_0$ and $g_1$, of size $2500$ each. First, we generate base scores for group 0  given by $q^{(g_0)}_a = -8 + \frac{8a}{100}$ for $a=0,...,99$. For group 1 we instead have $q^{(g_1)}_a = 8 - \frac{8a}{100}$. We then compute the score for user $u$ and candidate $a$ by sampling $z_{u,z}\sim \mathcal{N(0,\sigma^2)}$ and setting $q_{u,a} = I_{g_1}(u)q^{(g_1)}_a + I_{g_0}q^{(g_0)}_a + z_{u,a} $. The strong polarization data is generated with $\sigma=0.5$, while for the weak polarization data $\sigma=3$

\newpage 

\subsection{Additional real-data experiment results}
\label{appendix:amazon_books_demo_combined_gem}

\begin{figure}
    \centering
    \includegraphics[width=0.6\linewidth]{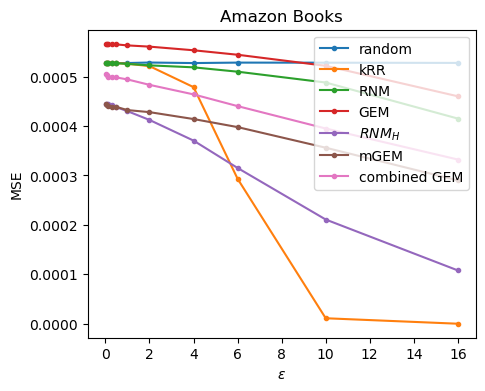}
    \caption{Evaluating the use of a portion of the privacy budget to automatically choose a selection mechanism on the Amazon books dataset.}
    \label{fig:amazon_books_demo_combined_gem}
\end{figure}

We briefly demonstrate that on the Amazon Books dataset, it is not productive to spend privacy budget to automatically choose a selection mechanism, see Figure \ref{fig:amazon_books_demo_combined_gem}.

\subsection{Synthetic Bandit Simulation with Distribution Shift}
\label{appendix:distr_shift}

\subsubsection{Experimental Design}
 At each iteration of the bandit simulation the learner picks one of two arms, $a=0,1$, and observes a clipped reward $r_{a,t} \sim \mathcal{N}(m_a,\sigma_a^2)$, where mean and variance differ between the arms. The means and variances of both arms are swapped at time $t=t_{shift}$ to simulate label distribution shift.

\paragraph{Non-private baseline}
As a non-private baseline we use the well known Upper-Confidence Bound algorithm (UCB) which computes an optimistic  score for each arm and selects the arm with maximum score. In particular the predicted score for arm $a$ at time $t$ is given by 
$$\hat{q}_{a,t} = \frac{1}{N_t(a)} \sum_{t'=1}^t r_{t'} I(a_{t'}=a) + \alpha \sqrt{\frac{\ln(T)}{N_t(a)}},$$
where $N_t(a)$ is the number of times action $a$ was chosen up to time $t$ and $\alpha$ is a hyper-parameter determining the level of optimism, commonly set to $\sqrt{2}$. Intuitively, the optimism means that UCB will be more likely to choose actions which haven't been chosen a lot in the past.

\subsubsection{Differentially-private algorithms}
We have two privacy requirements: 1) that the learner be differentially private, since it is shared between users and could leak information about observed rewards, and 2) that the chosen actions be private since they will be recorded precisely at the server. 

Since the learners rely on the empirical mean of rewards observed in the past, we can achieve requirement 1 by differentially-privately estimating the mean $\hat{q}_{a,t} = \frac{1}{N_t(a)} \sum_{t'=1}^t r_{t'} I(a_{t'}=a)$ for each action $a$ (note that this is a pure  bandit setting in which the reward does not depend on a context, but only on the chosen action). This is achieved with a tree-based mechanism \cite{10.1145/1806689.1806787}. In particular, if the chosen actions are DP (requirement 2), one can condition on the actions and run a tree-based mechanism for each action as well as assume the count $N_t(a)$ is already protected. Consequently, the tree-based mechanism only noises $\sum_{t'=1}^t r_{t'} I(a_{t'}=a)$. Finally, since even if the final time horizon $T$ of the bandit is known, the final count for each action $N_T(a)$ is unknown. Hence we use the hybrid mechanism introduced in \cite{chan2010PrivateContinual}, which combines the Logarithmic and binary mechanisms for unknown time horizons. The Hybrid mechanism ensures the released sequences $(\hat{q}_{a,t}\; for \; t\in \{t'\in\mathbb{N}\; : t'\leq T \,\&\, a_{t'}=a\})$ for $a=0,1$ are $\epsilon_M$-DP.

To achieve requirement 2, we utilize the differentially private selection mechanisms discussed in detail in Section \ref{sub:policies}. Rather than selecting the action with the highest predicted score $a_t = \arg\max_a (\hat{q}_{a,t})$, we select $a_t = \mathcal{M}(\hat{q}_{a,t}, \hat{\Delta}_{a,t}),$ where $\mathcal{M}$ is a DP-selection mechanism and $\hat{\Delta}_{a,t}$ are differentially-privately estimated truncated sensitivities. Note that sensitivities are not required for the kRR mechanism and RNM does not rely on action-wise sensitivities but on the maximum $\underset{a}{\max}\{\hat{\Delta}_{a,t}\}$. The privacy parameter for the private selection mechanisms is set to $\epsilon_S$ ensuring each action recorded at time t is $\epsilon_S$-DP.

The arm specific truncated sensitivities, $\hat{\Delta}_{a,t}$, used for randomized selection with GEM and $\rsy$ are computed by obtaining the 10th and 90th percentiles, $p_a^{0.1}$ and $p_a^{0.9}$, of the sets $\{r_{a,t-w},...,\cdots,r_{a,t}\}$, where $w$ is the specified window width. The quantiles are estimated differentially-privately with the Exponential mechanism, and we then set $\Delta_a=|p_a^{0.9}-p_a^{0.1}|$. Importantly, the Exponential mechanism is run such that each observed reward only participates in the estimation once, i.e. new sensitivities for arm $a$ are estimated again once $w$ new observations have been added since the previous estimation round. The pair of estimated quantiles for each action are set to be $\epsilon_p$-DP, so for 2 actions we have a total cost of $2\epsilon_p$.

In summary: The sequences of all released mean estimates for each action are $\epsilon_M$-DP, each action $a_t$ is locally $\epsilon_S$-DP, and each pair of estimated quantiles ($p_a^{0.1}$ and $p_a^{0.9}$) is $\epsilon_p$-DP. In our experiments, we set $\epsilon_M=1, \epsilon_S=1, \epsilon_p=1$ (to be paid twice, once for each action), and for kRR we set $\epsilon_S = 4$ since it does not need this additional information. Sensitivities are estimated based on the most recent $200$ observations and computed as the differentially privately estimated $0.1$ and $0.9$ quantiles of the past observations. For $t\leq t_{shift}$ the means of the reward distributions are $m_1=0.2,m_2=0.8$, with sensitivities $\sigma_1=0.1, \sigma_2=0.3$, while for $t > t_{shift}$ they are set to $m'_1=0.8,m'_2=0.2$, and
$\sigma'_1=0.3, \sigma'_2=0.1$.

\subsection{Alternative measures of correlation}
\label{s:weighted_correlation}

We have demonstrated that the Spearman coefficient is generally well-aligned with relative algorithm performance; in cases of positive correlation between scores and sensitivities, mGEM and $\rsy$ perform best, while in cases of negative correlation original GEM does. We also observed that catastrophic failure is possible for all selection algorithms that attempt to take advantage of heterogeneous noise (recall, e.g., Figure \ref{fig:second_scenarios}), and we suggested that a safe choice when little is known about the data distribution is RNM. Because we found positive correlation to be most common in the real datasets we explored, we now introduce an additional tool for deciding when to use RNM or one of the positive correlation heterogeneous mechanisms, mGEM or $\rsy$.

\begin{figure}
    \centering
    \includegraphics[width=\linewidth]{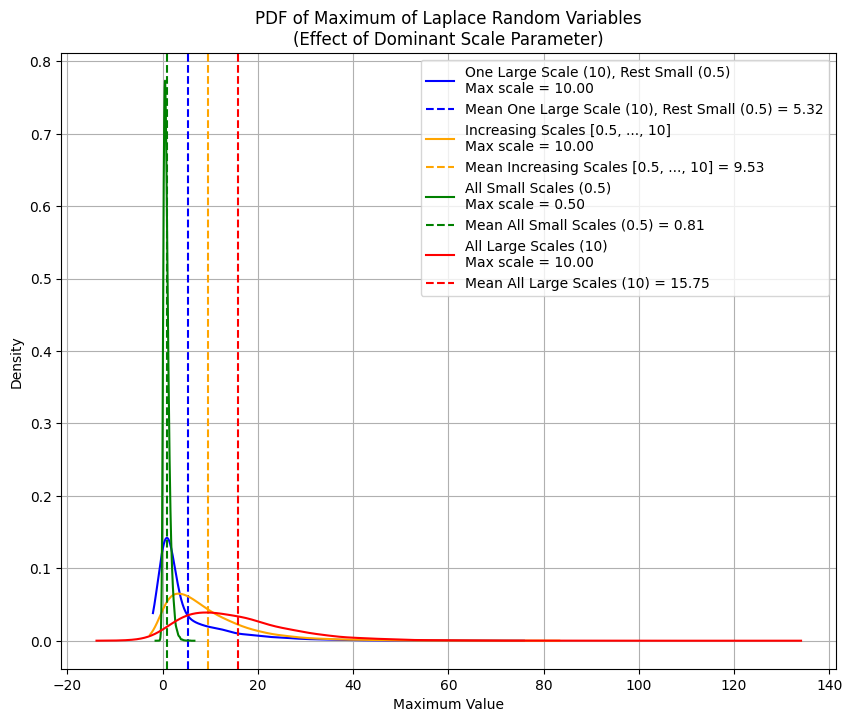}
    \caption{Estimated probability density function of the maximum of independent Laplace random variables with different scale parameters. In green, the scale parameters of all random variables are set to $0.5$. In blue, all but one scale parameters equal $0.5$ and the remaining one is set to $10$. In yellow, the scale parameters are linearly increasing from $0.5$ to $10$. In red, considers all scale parameters equal to $10$.}
    \label{fig:laplace_max}
\end{figure}

Consider a set of $A$ candidates with the same score but varying candidate-wise sensitivities $\Delta_a$. Then for RNMH the selected candidate is that with the highest noise sample. We study the case of Laplace noise where we have $z_a \sim {\rm Laplace}(b_a)$ with scale parameter $b_a \propto \Delta_a$. The CDF of $M = \underset{a}{\max}\{ z_a \}$ is $F_M(x)= \prod_{a}Pr(z_a\leq x)$, which for $x>0$ equates to 
\begin{equation}
    \prod_a \left( 1- \frac{1}{2} e^{-x/b_a}\right).
    \label{eq:lap_max}
\end{equation}
Note that $Pr(M\leq 0) = 2^{-A}$, which quickly becomes negligible for a larger number of candidates. From eqn~\eqref{eq:lap_max} it follows that the CDF of the maximum is predominantly determined by the candidates with larger scale parameters and equivalently larger sensitivities (see also Figure \ref{fig:laplace_max} for an empirical estimate of the pdfs; we observe that even one large scale parameter can significantly skew the distribution of the maximum). Splitting the range of scores into chunks, containing candidates with similar scores, it follows that probability of selecting a candidate from one of the buckets is predominantly influenced by the larger sensitivities of candidates in a given bucket. This motivates a weighted correlation measure, which intuitively assigns smaller weights to candidates with lower sensitivities than the maximum sensitivity of candidates in the same bucket, to indicate when heterogeneous noise is beneficial.

Formally, let the input dataset  $D$  consist of  $A$ candidates $\{(q_a, \Delta_a)\}_{a=1}^{A}$, where $ q_a $ represents the score for the $a$-th candidate, $\Delta_a $ represents the candidate-wise sensitivity of the \( a \)-th candidate.

The hyper-parameter $B$ represents the number of buckets or partitions of the range of scores, which we set to $5$ in our evaluation. Given the minimum and maximum values of the scores $q_{\text{min}} = \min(q_1, q_2, \dots, q_A)$ and $\quad q_{\text{max}} = \max(q_1, q_2, \dots, q_A)$, the range of scores $[q_{\text{min}}, q_{\text{max}}]$ is divided into $B$ buckets. Each bucket $b$ contains candidates with scores falling into the half-open interval $I_b = \left[ q_{\text{min}} + (b-1)\frac{q_{\text{max}} - q_{\text{min}}}{B}, \; q_{\text{min}} + b\frac{q_{\text{max}} - q_{\text{min}}}{B} \right)
$
for $b = 1, 2, \dots, B-1$ and $I_B$ is the closed interval.

For each bucket $b$ we compute the maximum sensitivity of candidates in that bucket as $\max \{ \Delta_a \;|\; a \; :\; q_a \in I_b  \}$. We then compute the weight of candidate $a$ in bucket $b$ by scaling the candidate's sensitivity by the maximum sensitivity in its bucket, which yields
$$w_a = \frac{\Delta_a}{\max \{ \Delta_a \;|\; a \; :\; q_a \in I_b  \}}.$$

Having defined the weights we now compute the weighted correlation. Let \( \mathbf{q} = (q_1, q_2, \dots, q_A) \), \( \mathbf{\Delta} = (\Delta_1, \Delta_2, \dots, \Delta_A) \), and \( \mathbf{w} = (w_1, w_2, \dots, w_A) \). We compute the weighted correlation as 

\begin{equation}
    \rho_w(\mathbf{q}, \boldsymbol{\Delta}) = \frac{\sum_{a=1}^{A} w_a (q_a - \mu_q)(\Delta_a - \mu_{\Delta})}{\sqrt{\sum_{a=1}^{A} w_a (q_a - \mu_q)^2} \cdot \sqrt{\sum_{a=1}^{A} w_a (\Delta_a - \mu_{\Delta})^2}},
    \label{eq:weighted_corr}
\end{equation}
where $\mu_q$ and $\mu_{\Delta}$ are the weighted means of scores and candidate-wise sensitivities, respectively.

\begin{figure}[t]
        \centering
  \begin{subfigure}[b]{0.5\textwidth}
            \begin{subfigure}[b]{0.49\textwidth}
                \centering
                \captionsetup{justification=centering}
                \includegraphics[width=.9\textwidth, trim={0 0 0 .71cm},clip]{plots/correlation/down_1.png}
                \caption{Scenario 1: \\Pearson Correlation: $0.22$\\
                        Spearman Correlation: $0.12$ \\ Weighted correlation: $0.83$}
            \end{subfigure}
            \begin{subfigure}[b]{0.49\textwidth}
                \centering
                \captionsetup{justification=centering}
                \includegraphics[width=.9\textwidth, trim={0 0 0 .71cm},clip]{plots/correlation/skewed_44.png}
                \caption{Scenario 2: \\Pearson Correlation: $0.44$\\
                        Spearman Correlation: $0.27$ \\ Weighted correlation: $0.16$}
            \end{subfigure}
        \end{subfigure}
        \begin{subfigure}[b]{.5\textwidth}
            \begin{subfigure}[b]{0.49\textwidth}
                \centering
                \captionsetup{justification=centering}
                \includegraphics[width=.9\textwidth,clip]{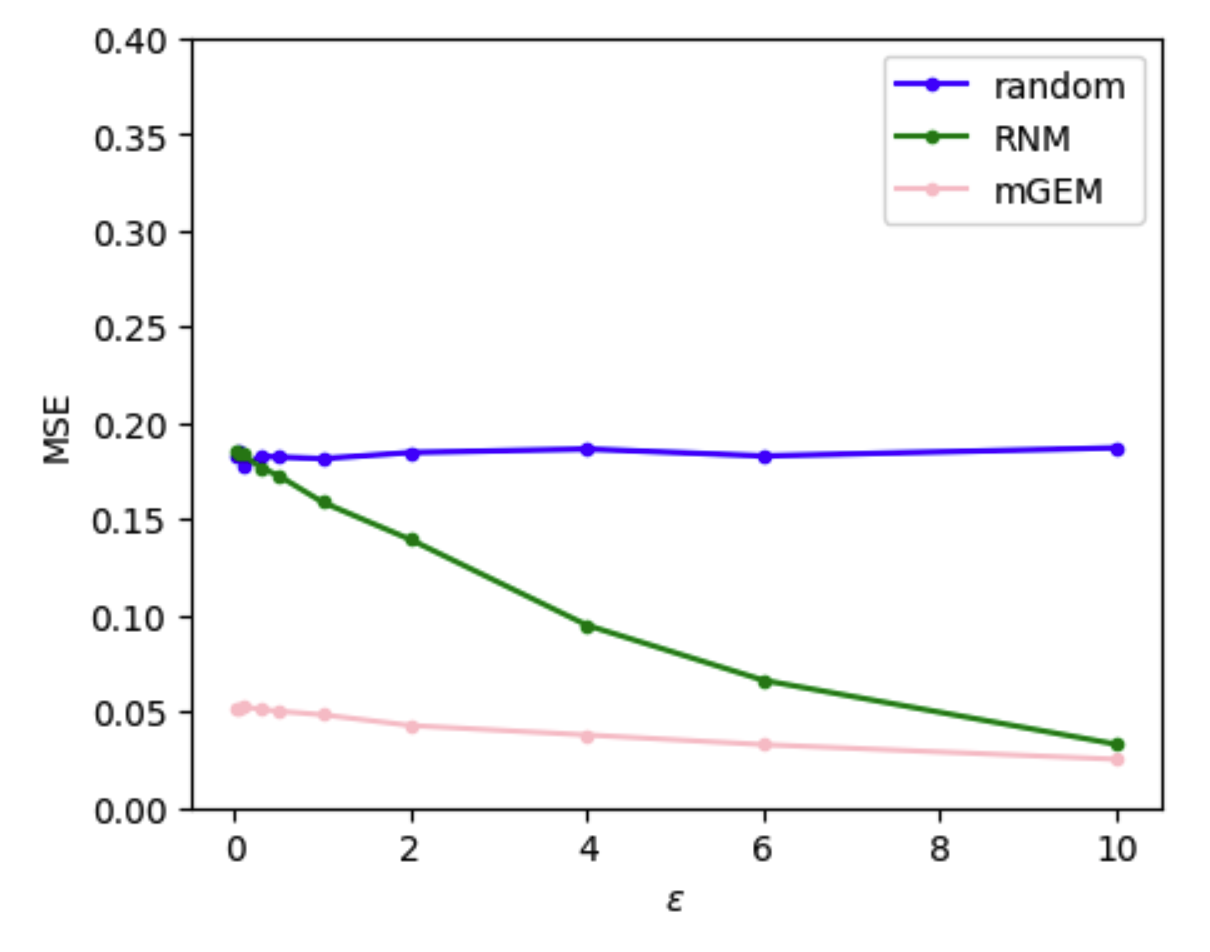}
                \caption{Performance of select DP selection mechanisms on Scenario 1}
            \end{subfigure}
            \begin{subfigure}[b]{0.49\textwidth}
                \centering
                \captionsetup{justification=centering}
                \includegraphics[width=.9\textwidth,clip]{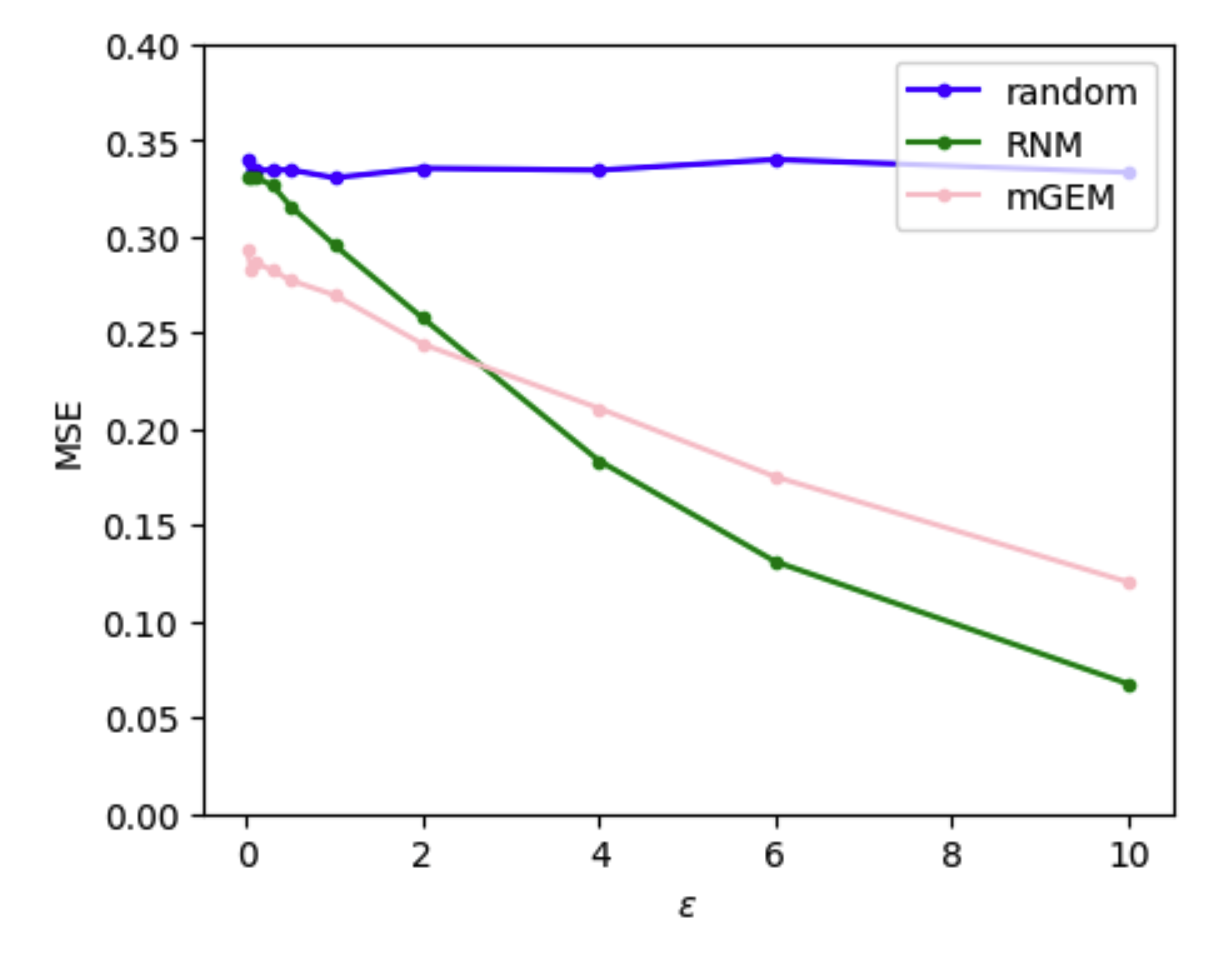}
                \caption{Performance of select DP selection mechanisms on Scenario 2}
            \end{subfigure}
    \end{subfigure}
        
    \caption{The top row show the sensitivities plotted against scores for the same distributions introduced in Fig. \ref{fig:correlation_scenarios}. The second row show mechanism performance in terms of MSE at different values of the privacy parameter $\epsilon$ for a subset of the algorithms discussed earlier. }
    \label{fig:correlation_scenarios_less}
\end{figure}

Returning to our synthetic data distributions from Figure \ref{fig:correlation_scenarios}, we observe (Figure~\ref{fig:correlation_scenarios_less}) that this weighted correlation metrics more accurately reflects the behavior of mGEM relative to RNM compared to Pearson or Spearman correlation coefficients. 

\subsection{Utility bound analysis}
\label{a:ubounds}

We find that GEM will have a utility bound worse than RNM when the sensitivity of the best candidate is greater than half the maximum sensitivity of all candidates, i.e.:

\begin{equation}
\Delta_* > \frac{\max_i \Delta_i}{2} 
\end{equation}

 Where $\Delta_*$ is the sensitivity of the highest score candidate. Furthermore, GEM will have a utility bound worse than selecting a candidate uniformly at random when the following condition is met:

\begin{equation}
\Delta_* > \frac{(q_*-q_-)}{4\frac{\log\actionspace/\beta}{\epsilon}}
\end{equation}

Where $q_*$ and $q_-$ are the scores of the best and worst candidates, respectively. When the best candidate's sensitivity is large relative to the range of all possible scores, GEM's worst-case utility degrades. This can also be visualized as the lower bound of the probability density of selecting the best candidate approaching the worst candidate; lower values of $\epsilon$ and $\beta$, and higher values of $\Delta_*$ and $\actionspace$ will decrease the lower bound. 

We propose that these utility bounds might also be used as heuristics to decide which algorithm is most suitable to any particular use case.

\end{document}